\documentclass[fleqn,usenatbib]{mnras}

\usepackage{newtxtext,newtxmath}

\usepackage[T1]{fontenc}

\DeclareRobustCommand{\VAN}[3]{#2}
\let\VANthebibliography\thebibliography
\def\thebibliography{\DeclareRobustCommand{\VAN}[3]{##3}\VANthebibliography}

\usepackage{graphicx}	
\usepackage{amsmath}	

\pdfoutput=1
\usepackage[T1]{fontenc}
\usepackage{graphicx}
\usepackage{wasysym}
\usepackage{esint}
\usepackage[authoryear]{natbib}
\usepackage[utf8]{inputenc}
\usepackage{threeparttable}
\usepackage{hyperref}
\usepackage{xspace}
\usepackage{url}
\usepackage{babel}
\usepackage{ulem}
\usepackage{newtxtext,newtxmath}
\usepackage{hyperref}	
\usepackage{mathtools}
\usepackage{xcolor}
\usepackage{multirow}
\usepackage{orcidlink}
\usepackage{upgreek}

\hypersetup{colorlinks=true,linkcolor=blue,citecolor=blue,filecolor=blue,urlcolor=blue,}

\newcommand{\arepo}{\textsc{Arepo}\xspace}

\newcommand{\crayon}{\textsc{Crayon+}\xspace}
\newcommand{\crest}{\textsc{Crest}\xspace}

\title[Revisiting radio synchrotron diagnostics]{Revisiting radio synchrotron diagnostics in star-forming galaxies}
\author[M. Werhahn et al.]{Maria Werhahn$^{1,}$\thanks{E-mail:
mwerhahn@mpa-garching.mpg.de}\orcidlink{0000-0003-4984-4389},
Christoph Pfrommer$^2$\orcidlink{0000-0002-7275-3998}, 
Philipp Girichidis$^3$\orcidlink{0000-0002-9300-9914}, 
Joseph Whittingham$^2$\orcidlink{0000-0002-0198-8490},
L\'{e}na Jlassi$^2$\orcidlink{0009-0007-9039-294X},
\newauthor R\"udiger Pakmor$^1$\orcidlink{0000-0003-3308-2420},
Rebekka Bieri$^4$\orcidlink{0000-0002-4554-4488},
Rainer Weinberger$^2$\orcidlink{0000-0001-6260-9709},
Volker Springel$^1$\orcidlink{0000-0001-5976-4599},
Freeke van~de~Voort$^5$%
\orcidlink{0000-0002-6301-638X}
\vspace*{0.1cm}\\%
$^{1}$Max-Planck-Institut f\"ur Astrophysik (MPA), Karl-Schwarzschild-Str. 1, 85748 Garching, Germany\\%
$^{2}$Leibniz-Institut f\"ur Astrophysik Potsdam (AIP), An der Sternwarte 16, 14482 Potsdam, Germany\\%
$^{3}$Universit\"{a}t Heidelberg, Zentrum f\"{u}r Astronomie, Institut f\"{u}r Theoretische Astrophysik, Albert-Ueberle-Str. 2, 69120 Heidelberg, Germany\\%
$^{4}$University of Zurich, Department of Astrophysics, Winterthurerstrasse 190, 8057 Zurich, Switzerland\\%
$^{5}$Cardiff Hub for Astrophysics Research and Technology, School of Physics and Astronomy, Cardiff University, Queen’s Buildings, Cardiff CF24 3AA, UK
}

\date{Accepted XXX. Received YYY; in original form ZZZ}

\pubyear{2026}

\begin{document}
\label{firstpage}
\pagerange{\pageref{firstpage}--\pageref{lastpage}}
\maketitle

\begin{abstract} 
Radio continuum observations are widely used to study cosmic ray (CR) electron populations and transport processes in star-forming galaxies, but their interpretation relies on several simplifying assumptions. Here, we revisit three common assumptions: that some vertical radio profiles can be explained by CR advection alone, that radio spectra directly trace the galaxy-wide CR electron spectrum, and that bremsstrahlung and Coulomb losses are negligible for radio-emitting electrons.
We model radio emission using time-dependent CR electron evolution in a magnetohydrodynamical simulation of an isolated Milky Way–mass galaxy. CR electron spectra are evolved self-consistently along Lagrangian tracer particles with the \crest framework, including injection at supernova remnants, advection with the gas, and spatially and temporally varying radiative losses. We compare these results to commonly adopted steady-state models.
We find that advection-only transport in self-consistently driven galactic winds fails to reproduce the extended vertical radio intensity profiles observed in edge-on galaxies, despite reproducing the observed steepening of spectral indices with height. This is because slowly accelerating winds keep electrons in strong cooling environments for too long. Matching observed radio haloes with advection alone requires unrealistically high midplane wind velocities, implying that additional transport or re-acceleration processes are required.
Although galaxy-integrated CR electron spectra at radio-emitting energies are similar across models, the resulting synchrotron spectra differ systematically because radio emission is biased toward young electrons in dense, strongly magnetised regions. Finally, we show that bremsstrahlung and Coulomb losses significantly shape radio spectra even when their loss rate is subdominant and therefore cannot be neglected.
\end{abstract}




\section{Introduction}\label{sec:introduction}

Radio continuum emission from star-forming galaxies is widely used not only as a tracer of star formation activity, but also as a tracer of cosmic ray (CR) electron physics. Synchrotron radiation probes the interaction of CR electrons with magnetic fields, and could potentially give us a window to trace galactic scale outflows, both from purely star-formation driven outflows \citep[e.g.\ ][]{2009Heesen, 2020Stein, 2024Heesen, 2025Koribalski, 2025Matthews}, as well as outflows related to active galactic nuclei \citep[e.g.\ ][]{1998Elmouttie, 2003IrwinSaikia, 2025Veronese}. Fortunately, present-day radio observations provide a wealth of data, ranging from integrated radio spectra to spatially resolved maps of radio haloes, including vertical intensity and spectral index profiles for edge-on galaxies.

Understanding the physical origin of these radio observables is of particular importance because constraints on CR electron transport can inform our broader picture of CR feedback in galaxies. While CR protons dominate the CR energy budget and are therefore dynamically most relevant, they are observationally far less accessible: their hadronic interactions produce gamma-ray emission, which has been robustly detected only for the Milky Way and a small number of nearby star-forming galaxies \citep{2009VERITAS_M82, Abdalla2018HESS, 2022FermiLAT}. In contrast, radio observations of CR electrons are available for large samples of galaxies and at high spatial resolution, making them an attractive probe of CR physics, provided that their interpretation is well understood.

A variety of modelling approaches have been employed to interpret radio observations under various simplifying assumptions. One-zone models treat galaxies as spatially averaged systems in steady state, with gas density, magnetic field strength, and CR confinement times parametrized to match global observables such as radio spectra and the far-infrared–radio correlation \citep{2010Lacki, 2011Lacki, 2013Yoast-Hull, 2016Eichmann}. One-dimensional transport models have been used extensively to interpret radio haloes of edge-on galaxies, adopting simplified transport prescriptions, either advection or diffusion, and fitting the corresponding parameters to match observed vertical profiles \citep[e.g.\ ][]{2016Heesen, 2018Heesen, 2019Schmidt, 2019Stein, 2022Heald, 2023Stein, 2025Xu}.

More recently, numerical simulations have begun to incorporate increasingly detailed treatments of CR electron transport and synchrotron emission. Spectrally resolved CR electron transport has been implemented and tested in stratified-box simulations with the \textsc{Piernik} code \citep{2021Ogrodnik, 2025Baldacchino-Jordan}. A global application to the edge-on galaxy NGC~891 was presented by \citet{2023Ogrodnik}, who compared synthetic radio spectral index profiles to observations at low frequencies. However, this study did not consider bremsstrahlung and Coulomb losses, nor vertical radio intensity profiles, both of which we show here to be important for interpreting radio diagnostics.
Post-processing studies of synchrotron emission and Faraday rotation in stratified-box simulations of kpc-sized patches representative of the solar neighbourhood have been carried out in the \textsc{Silcc} project using semi-analytical CR models \citep{2022Rappaz}. 
More recently, \textsc{Tigress} simulations have modelled CR transport and synthetic synchrotron emission in a similar stratified-box setup. They employed a more self-consistent treatment of spatial CR transport, but without explicit transport in energy space and while still lacking coupling to the global evolution of a galaxy \citep{2025Linzer}.
Within the FIRE framework, CR-MHD simulations evolve CR proton and electron spectra self-consistently in Milky Way–mass zoom simulations \citep{2022Hopkins}. These have been used to investigate the equipartition assumption \citep{2024aPonnadaSpectrallyResolved} and the impact of different CR proton transport models on radio emission \citep{2024bPonnada}, but adopt fixed proton-to-electron ratios and simplified electron spectral shapes, and have not yet explored radio spectra or vertical intensity and spectral index profiles in detail.

In earlier work, we applied a cell-based steady-state model using the \crayon code to isolated galaxy simulations, successfully reproducing radio spectra and the far-infrared-radio correlation \citep{2021WerhahnIII, 2022Pfrommer}. This approach was extended by \citet{2024ChiuSandy} to simulations with a multiphase interstellar medium (ISM) and a two-moment CR transport prescription \citep{2025aThomas}. This reproduces several radio observables of NGC~4217, including its radio spectrum, vertical intensity profile, and X-shaped morphology indicative of a galactic outflow \citep{2020Stein} and allowed to assess the robustness of the magnetic equipartition assumption commonly used in radio synchrotron observations \citep{2025Chiu_equipartition}. However, because CR electron aging was not accounted for, spectral index profiles could not be studied within this framework.

Taken together, these studies suggest that a range of physical processes, such as the global galaxy evolution, CR injection and cooling histories, time-dependent spectral evolution, and spatial transport by advection, diffusion, and/or streaming, may all be important for correctly interpreting radio diagnostics. The relative importance of these processes, however, remains poorly constrained.

In this work, we aim to bridge several of these approaches. We apply the \crest (Cosmic Ray Electron Spectra that are evolved in Time) framework \citep{2019Winner} to model the time-dependent, spectrally-resolved evolution of CR electrons in a full MHD simulation. This has previously been successfully applied to simulating the multi-frequency emission of the supernova (SN) remnant SN1006 \citep{2020Winner}, the physics underlying radio relics in galaxy clusters \citep{2026aWhittingham, 2026bWhittingham}, and radio emission from AGN jets in an isolated Perseus-like galaxy cluster \citep{2026Jlassi}. Here, we apply it to an MHD simulation of a star-forming Milky Way-like galaxy and compute the resulting radio synchrotron emission. By following Lagrangian tracer particles advected with the gas, this approach allows us to isolate the effects of advection on CR electron spectra and radio observables. 
In a companion paper, we applied \crest for the first time to global galaxy simulations and analysed the resulting CR electron spectra in detail, comparing them to steady-state models and to local CR measurements \citep{2025Werhahn_CREST}.
Building on this, we here focus on the observable radio emission and compare the simulated radio spectra, vertical intensity profiles, and spectral index profiles to predictions from one-zone and cell-based steady-state models, confronting them with observational constraints. 
This allows us to move from the intrinsic CR electron population to the radio diagnostics used in observations, and to assess how robust common interpretations of radio data really are.

In particular, we aim to test three questions that are widely assumed to be true in the literature:
\begin{itemize}
    \item Can the vertical radio profiles from star-forming galaxies be explained by purely advective CRs?
    \item Does the radio spectrum trace the underlying, galaxy-wide CR electron spectrum?
    \item Are bremsstrahlung and Coulomb losses negligible for radio-emitting electrons?
\end{itemize}

This paper is organised as follows. In Section~\ref{Sec:Simulations}, we describe the galaxy simulation, the time-dependent CR electron modelling with \crest, the steady-state framework \crayon, and the calculation of radio synchrotron emission. Section~\ref{sec:GlobalRadioEmission} presents a global overview of the simulated radio emission and compares the different modelling approaches to observations of NGC~891. 
In Section~\ref{sec:VerticalProfilesAdvection}, we test whether advection-only transport can reproduce observed vertical radio profiles in edge-on galaxies. Section~\ref{Sec:GlobalRadioGlobalElectronSpectra} examines the connection between CR electron spectra and radio synchrotron spectra, while Section~\ref{sec:BremsstrahlungCoulombLosses} discusses the role of bremsstrahlung and Coulomb losses. We discuss the impact of free-free emission and absorption in Section~\ref{sec:ff-emission}, the implications of our results in Section~\ref{sec:DiscussionVerticalProfiles}, and summarise our conclusions in Section~\ref{sec:conclusions}.

\section{Simulations and methods}
\label{Sec:Simulations}

Our analysis is based on a simulation of an isolated Milky Way–mass galaxy performed with the moving-mesh MHD code \arepo \citep{2010Springel,2016cPakmor,2020Weinberger}. The simulation setup is identical to that used in our companion paper \citep{2025Werhahn_CREST}, where we presented the first application of \crest to a global galaxy simulation and analysed the resulting CR electron spectra in detail. We therefore only summarise the key aspects here and refer the reader to that work for a full description of the numerical methods.

We initialise an isolated gas cloud embedded in a static dark matter halo with an NFW profile of mass $M_{200}=10^{12}\,\mathrm{M_\odot}$ and concentration $c_{200}=7$. The simulation is initialised with $10^7$ gas cells, each with a target mass of $1.55\times10^4\,\mathrm{M_\odot}$. This target mass is maintained to within a factor of two through adaptive mesh refinement and derefinement. Radiative cooling and star formation are modelled using the effective equation-of-state prescription of \citet{2003SpringelHernquist}. The system naturally evolves into a rotationally supported disc galaxy with a stellar mass and star formation rate (SFR) comparable to those of the Milky Way. After an initial starburst phase, the SFR declines steadily to a few $\mathrm{M_\odot\,yr^{-1}}$ after a few Gyr.
We evolve the equations of ideal MHD, starting from a uniform seed magnetic field of strength $10^{-10}$~G, which is rapidly amplified by compression as well as by a turbulent dynamo to saturate close to equipartition with the kinetic turbulence within the disc \citep{2022Pfrommer}. CRs are injected at SN sites, with a fixed fraction $\zeta_\mathrm{SN}=0.1$ of the SN energy converted into CR energy and distributed to neighbouring gas cells. CR transport is treated using a one-moment diffusion–advection scheme with a constant diffusion coefficient $\kappa_\parallel=10^{28}\,\mathrm{cm^2\,s^{-1}}$ along magnetic field lines, which drives the formation of large-scale, magnetised galactic outflows.
Our fiducial run employs super-Lagrangian refinement in the central regions to better resolve the disc–halo interface and the emerging winds. Specifically, we apply a nested refinement scheme following \citet{2025aThomas}, defining four concentric spherical shells around the centre of the simulation with radii 
$r_\mathrm{shell}=\{15,30,60,120\}$~kpc. Within each shell, a maximum cell size of $\Delta x_\mathrm{max}=\{100,200,400,1000\}~$pc is enforced by refining gas cells whose volumes exceed $V>(4\pi/3)\Delta x_\mathrm{max}^3$. This ensures enhanced spatial resolution in the central regions and along the developing galactic outflows.

\subsection{Time-dependent electron modelling with \crest}

Because CR electrons contribute negligibly to the total CR energy density, they do not affect the dynamical evolution of the simulated galaxy. This allows us to compute their spectral evolution in post-processing using the \crest framework, introduced in \citet{2019Winner}, following Lagrangian tracer particles that are advected with the gas. In addition to sampling local gas properties, these tracers carry CR electron energy spectra that are injected at SN sites and evolved in time, providing a Lagrangian description of the CR electron population. We then solve the one‑dimensional Fokker–Planck equation in momentum space along each trajectory to evolve the electron spectra, accounting for adiabatic compression or expansion, losses due to Coulomb interactions, as well as bremsstrahlung, inverse Compton (IC) and synchrotron losses.
Gas density and magnetic field strength are taken directly from the host cell of each tracer. The IC cooling rate is computed using a model for the ambient radiation field that is consistent with our steady-state post-processing approach: the photon energy density is given by the sum of the CMB and a locally varying interstellar radiation field estimated from the star formation activity in the galaxy. The latter is derived from the far-infrared luminosity associated with star-forming cells, following \citet{1998Kennicutt}.

Tracer particles are introduced within a cylinder ($R=30$~kpc, $|{z}|<10$~kpc) after a rotationally supported disc has formed, ensuring that they reliably follow the gas flow, and 500~Myr before the time of analysis, which is sufficient to recover the evolution history of the spectra due to short electron cooling timescales \citep{2025Werhahn_CREST}. Additionally, we spawn tracers in the central region of the galaxy (i.e. a cylinder of radius 5~kpc and total height of 20~kpc) in cells which are devoid of tracers to improve the sampling of galactic outflows and ensure that each gas cell retains at least one tracer. At the time of analysis, this results in a total of approximately $10^7$ tracer particles.
We inject energy into CR electrons (represented by the tracer particles) at SN sites with a power-law source spectrum of slope $\alpha_\mathrm{inj}=2.2$. This is normalised such that a fixed fraction of the CR proton energy injected by SNe is transferred to electrons, i.e.\ $\zeta_\mathrm{ep}=0.22$. 
This choice reproduces the locally observed CR electron‑to‑proton ratio of 0.01 at 10~GeV at the solar radius \citep{2025Werhahn_CREST}. In addition, it yields radio luminosities consistent with the observed correlation between radio luminosity and SFR (see Fig~\ref{fig:FRC}).
The spectrum $f(p)$ is evolved over a wide momentum range, from $p=10^{-1}$ to $10^8$. Here, we define the normalised electron momentum $p$ in units of the electron rest mass $m_\mathrm{e}$ times the speed of light $c$, i.e. $p=P/(m_\mathrm{e}c)$. We sample the spectrum with 20 bins per decade, and include a physically motivated high-energy cut-off at an electron energy of 20~TeV. 
A detailed description of the injection model, spectral normalisation, and numerical implementation is given in our companion paper \citep{2025Werhahn_CREST}, to which we refer the reader for further details.

\subsection{Steady-state CR electron modelling with \crayon}
To benchmark the time-dependent \crest results, we compute steady-state CR electron spectra using the cell-based post-processing framework \crayon \citep{2021WerhahnI}. In this approach, the CR electron spectrum in each simulation cell is obtained by solving the steady-state diffusion-loss equation, which balances continuous injection against energy losses, as well as escape processes. The loss terms include synchrotron, IC, bremsstrahlung, and Coulomb cooling, evaluated using the local gas density, magnetic field strength, and radiation field, consistent with the \crest modelling.
CR electrons are injected with the same functional form and spectral parameters as adopted in the time-dependent modelling, enabling a direct comparison between the two approaches. Escape losses are parameterised through effective advection and diffusion timescales. Since spatial diffusion is not included for the tracer particles and can therefore not be captured by \crest in post-processing, we run \crayon both in its standard configuration and in a variant where the parameterized diffusion in the steady-state equation is suppressed. This allows us to isolate the impact of this assumption on the resulting steady-state spectra.
Unless stated otherwise, comparisons between \crest and the steady‑state approach are performed using the \crayon model without diffusion.

The normalisation of the steady-state CR electron spectra is set by requiring consistency with local measurements of the CR electron-to-proton ratio at 10~GeV. This calibration determines the injected electron-to-proton ratio in the steady-state models and serves as the reference point for the comparison with \crest. 
For the comparisons presented in this work, we adopt a fiducial normalisation in \crest that matches the steady-state model without diffusion. Since the underlying transport and loss physics is independent of this choice of the injected CR electron normalisation, the resulting spectral shapes are unaffected and can be rescaled in post-processing if required. 
Within our adopted choice of normalisation, however, both modelling approaches already reproduce the observed FIR–radio correlation (see App.~\ref{app:FRC}).

\subsection{Radio synchrotron emission}
We calculate the radio synchrotron emissivity with the \crayon code, following \citet{1986rpa..book.....R} as
\begin{align}
j_{\nu}=\frac{\sqrt{3}e^{3}B_{\perp}}{m_{e}c^{2}}\intop_{0}^{\infty}f(p)F(\nu/\nu_\mathrm{c})\mathrm{d}p,
\label{eq:j_nu_sync}
\end{align}
where $e$ denotes the elementary charge and $B_\perp$ is the magnetic field strength of the component perpendicular to the line of sight. We employ an analytical approximation for the synchrotron kernel $F(\nu/\nu_c)$ provided by \citet{2010PhRvD..82d3002A} to speed up the calculation \citep[for more details, see Appendix A1 in ][]{2021WerhahnIII}.
The typical frequency of radio synchrotron emission from a CR electron with Lorentz factor $\gamma_\mathrm{e}$ is given by \citep[see e.g.\ ][]{2021WerhahnIII}
\begin{align}
    \nu_\mathrm{syn}=\frac{3eB_\perp}{2\pi m_\mathrm{e}c}\gamma_\mathrm{e}^2 \simeq 1~\mathrm{GHz} \frac{B_\perp}{1~\upmu \mathrm{G}} \left(\frac{\gamma_\mathrm{e}}{10^4}\right)^2.
\end{align}
Assuming $p\gg1$, the corresponding typical electron momentum emitting into the 1.4~GHz band can be estimated via 
\begin{align}
    p|_\mathrm{1.4~GHz}\simeq 1.3\times10^4\times \left(\frac{B_\perp}{1~\upmu \mathrm{G}}\right)^{-1/2}.
    \label{eq:p_syn}
\end{align}
Furthermore, we follow the notation used in \citet{2021WerhahnIII} and define the spectral intensity and luminosity, in the absence of any absorption or other emission process, as the integral along the line of sight $s$ as
\begin{align}
    I_\nu = \frac{1}{4\pi }\int j_\nu \mathrm{d}s
\end{align}
and
\begin{align}
    L_\nu = \int j_\nu \mathrm{dV},
\end{align}
where $V$ denotes the integration volume.
We discuss the impact of including free-free absorption and emission in Section~\ref{sec:ff-emission}.
Furthermore, the spectral slope of the radio spectrum is defined such that the luminosity scales as $L_\nu \propto \nu^{-\alpha_\nu}$, i.e. via
\begin{align}
    \alpha_\nu = -\frac{\mathrm{d}\log L_\nu}{\mathrm{d}\log \nu}.
\end{align}

\section{Global radio emission properties of the simulated galaxy}\label{sec:GlobalRadioEmission}

Before turning to the three assumptions introduced in Section~\ref{sec:introduction}, we first provide a global overview of the radio emission properties predicted by the different modelling approaches.

\begin{figure}
    \centering
    \includegraphics[]{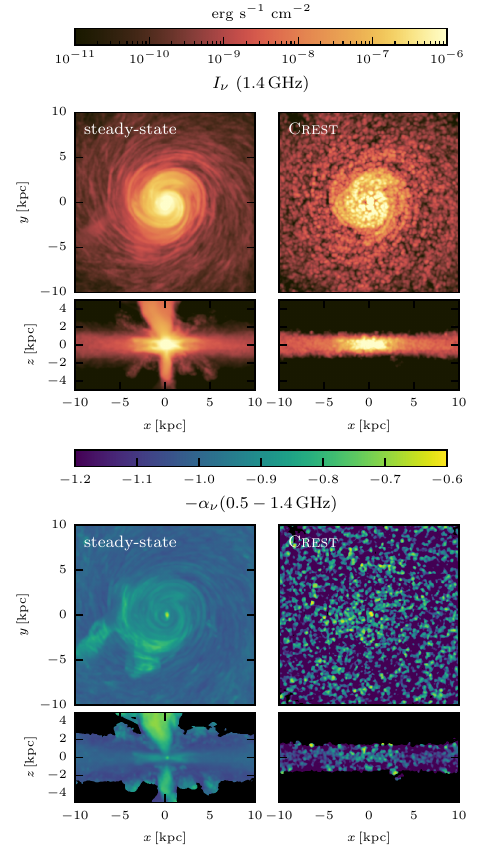}
    \caption{Face-on and edge-on radio intensity (upper four panels) and spectral index maps (lower four panels) at $t=1$~Gyr for the steady-state model obtained with \crayon (left-hand panels) and the time-dependent modelling with \crest (right-hand panels). 
    The radio intensity maps are smoothed with a Gaussian kernel of FWHM 0.2~kpc. The spectral index maps are computed from the smoothed intensity maps at 0.5 and 1.4~GHz, applying a mask that includes only pixels with intensities above $10^{-11}\,\mathrm{erg\,s^{-1}\,cm^{-2}}$.
    The steady-state model produces smoother and more vertically extended radio emission, while the \crest model exhibits a more compact and patchy intensity distribution and stronger local variations in the spectral index, reflecting the discrete nature of CR electron injection at SN sites.
    }
    \label{fig:map_emission_alpha}
\end{figure}

\subsection{Global radio emission maps}

Figure~\ref{fig:map_emission_alpha} presents face-on and edge-on radio intensity and spectral index maps at $t=1$~Gyr for the two modelling approaches considered in this work. This snapshot is representative of the global radio morphology throughout the evolution of the simulated galaxy. We note that the two post-processing models are based on the same underlying simulation and therefore the hydrodynamical quantities, such as the CR proton energy density, gas density and magnetic field properties, are identical.

In the face-on intensity maps, both models exhibit broadly similar large-scale emission patterns. The steady-state model produces a smoother and more spatially extended intensity distribution, whereas the \crest results appear more patchy, reflecting the discrete nature of CR electron injection and the explicit time-dependent evolution of individual electron populations. This patchiness is not driven by sampling noise: the emitting regions are well resolved by large numbers of tracer particles and spatially coincide with the young electron populations identified in Fig.~\ref{fig:map_position_histograms}.
Apart from this difference in small-scale structure, the face-on maps are qualitatively comparable.

More pronounced differences emerge in the edge-on intensity maps. In the cell-based steady-state approach, the local normalisation of CR electrons is tied to that of CR protons, which are transported relative to the gas via diffusion and cool more slowly. As a result, CR-driven outflows become magnetised and populated with CR protons, and the assumption of local injection--cooling equilibrium in each cell ensures a sufficient population of high-energy electrons to produce observable radio emission in the halo.

In contrast, the \crest model shows little to no detectable radio emission in the outflows. In this case, CR electrons are injected at SN sites and subsequently advected with the gas, while their spectral evolution is followed explicitly on Lagrangian tracer particles. High-energy electrons cool efficiently as they are carried away from the disc, so that they fall below the energies relevant for radio emission before reaching large heights above the midplane. This difference primarily reflects the moderate outflow velocities in our simulation, which are insufficient to transport radio-emitting electrons out of the disc before radiative losses dominate. We explore this point in more detail in Section~\ref{sec:VerticalProfilesAdvection}.

We now turn to the spectral index maps shown in the lower panels of Fig.~\ref{fig:map_emission_alpha}. As for the intensity maps, the steady-state results are noticeably smoother, both face-on and edge-on, whereas the \crest maps exhibit strong local variations in the spectral index. These variations directly trace the spatially and temporally varying CR electron spectra in the live modelling with \crest \citep{2025Werhahn_CREST}. Notably, regions of high radio intensity in the \crest maps tend to be associated with flatter spectral indices, indicating a stronger contribution from young, recently injected electron populations. We quantify this behaviour and its implications for interpreting observed radio spectra in Section~\ref{Sec:GlobalRadioGlobalElectronSpectra}.

\subsection{Global radio spectra and their time evolution}
\begin{figure}
    \centering
    \includegraphics[]{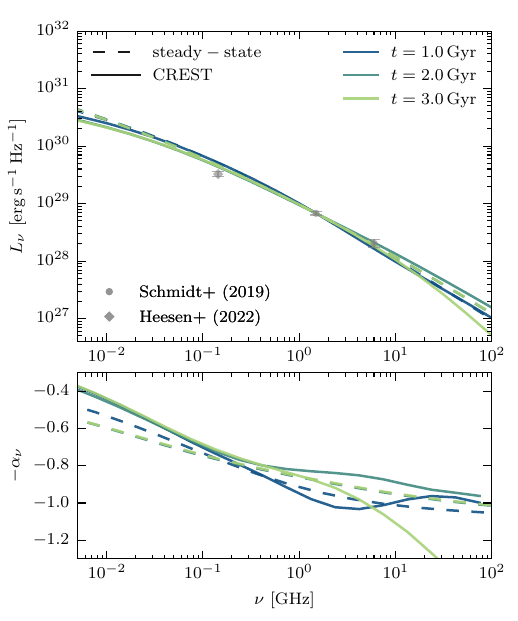} 
    \caption{Time evolution of the global edge-on radio synchrotron spectrum (top) and the corresponding spectral index as a function of frequency (bottom; defined as $L_\nu\propto\nu^{-\alpha{}}$) between $t=1$ and $3$~Gyr, for the time-dependent modelling with \crest as well as the steady-state model obtained with \crayon. The spectra are renormalised to the observed data of NGC~891 \citep{2019Schmidt} at 1.5~GHz but are consistent with the observed luminosity within factors of $<1.5$.
    At low frequencies, the \crest spectra are consistently flatter than those obtained from the steady-state model. At higher frequencies ($\nu \gtrsim 1$~GHz), the \crest spectra exhibit noticeable time variability, whereas the steady-state spectra remain largely unchanged.
    The observed radio spectrum from NGC~891 \citep{2019Schmidt, 2022Heesen} agrees well with the spectral shape at 1.5 and 6~GHz, while our models slightly over-predict the emission at 0.144~GHz.
    }
    \label{fig:radio_spectra}
\end{figure}

Figure~\ref{fig:radio_spectra} shows the global synchrotron spectra obtained with \crest and with the steady-state approach (\crayon) at three representative times, $t=1$, 2, and 3~Gyr. In addition to the spectra themselves, we display the corresponding spectral index as a function of observing frequency.
To allow for comparison of spectral shapes, the global spectra are renormalised to the same value at 1.5~GHz, i.e.\ the observed flux density $F_\nu$ of the well studied, edge-on Milky Way--like galaxy NGC~891 \citep[][]{2019Schmidt}, which has been extensively analysed in the radio. We include observed global radio fluxes at 144~MHz \citep{2022Heesen} and at 1.5 and 6~GHz \citep{2019Schmidt}, converting them to luminosities via $L_\nu=4\pi d^2 F_\nu$ assuming a distance of $d=9.1$~Mpc \citep{2011RadburnSmith}. 
The renormalisation factors at 1.5~GHz range from $0.7$–$1$ for the steady-state models and from $1.3$–$1.5$ for \crest, indicating that \crest systematically predicts lower luminosities despite adopting the same injected electron-to-proton ratio (see also Fig.~\ref{fig:FRC}).

\subsubsection{Comparison of the spectra obtained by the different models}
At low frequencies ($\nu\lesssim0.1$~GHz), the \crest spectra are systematically flatter than the steady-state spectra at all times. This difference is most pronounced at $t=3$~Gyr, where the steady-state model yields a spectral index of $\alpha_\nu\approx0.60$ (at 0.01~GHz), while the \crest spectra exhibit significantly flatter slopes of $\alpha_\nu\approx0.44$ (see lower panel of Fig.~\ref{fig:radio_spectra}).

At higher frequencies ($\nu\gtrsim2$~GHz), the synchrotron spectra obtained with \crest show clear temporal variability, with changes in spectral shapes appearing at different epochs. This behaviour reflects time-dependent variations due to injection and rapid cooling in the high-energy CR electron population, whose evolution is followed explicitly in the live modelling. By contrast, the steady-state spectra remain largely unchanged with time by construction, as they assume instantaneous equilibrium between injection and cooling.

Overall, while the global synchrotron spectra predicted by the \crest and steady-state models are broadly similar, systematic differences arise both in their detailed spectral shapes and in their temporal evolution. 
In the following sections, we examine the physical origin of these differences by revisiting several common assumptions underlying the interpretation of radio observations of star-forming galaxies.

\subsubsection{Comparison to NGC~891}
Comparing the simulated spectra to observations of NGC~891 in Fig.~\ref{fig:radio_spectra} shows that at frequencies of 1.5 and 6~GHz, the spectral shape is well reproduced within observational uncertainties by all models. At lower frequencies, the observed spectral flattening is not fully captured by any of the models. This discrepancy could potentially arise from free-free absorption \citep[see, e.g.,][]{2021WerhahnIII}, which can substantially flatten the spectrum below GHz frequencies, as we will discuss in Section~\ref{sec:ff-emission}.

Estimates of the SFR of NGC~891 range from $1.55~\mathrm{M_\odot\,yr^{-1}}$ \citep{2015Wiegert}, through $1.88~\mathrm{M_\odot\,yr^{-1}}$ \citep{2019Vargas}, up to $3.3~\mathrm{M_\odot\,yr^{-1}}$ \citep{2012Krause}.
The simulated galaxy exhibits higher SFRs than NGC~891, with values of $14.5$, $7.2$, and $4.4~\mathrm{M_\odot\,yr^{-1}}$ at $t=1$, 2, and 3~Gyr, respectively. Nevertheless, the total radio luminosity at 1.5~GHz agrees with observations within a factor of $\lesssim1.5$ at all times (see also Fig.~\ref{fig:FRC}).

The weak evolution of the radio luminosity despite the declining SFR is consistent with the behaviour of the global CR electron spectra found in \citet{2025Werhahn_CREST}. While the injection rate decreases with the SFR, the cooling rate is reduced in a similar manner, primarily due to the declining photon energy density that determines the IC losses. As a result, the normalisation of the electron spectra (and hence the radio luminosity) remains approximately constant over time.

Motivated by this overall agreement, we compare the vertical radio intensity and spectral index profiles of NGC~891 to the different modelling approaches in the following section.

\section{Can vertical radio profiles of star-forming galaxies be explained by purely advective CRs?}\label{sec:VerticalProfilesAdvection}

\begin{figure}
    \centering
    \includegraphics[scale=1]{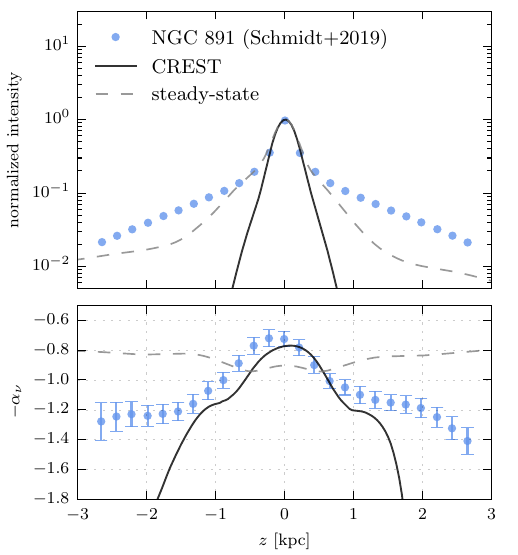}
    \caption{
    \textit{Upper panel:} Normalized vertical radio intensity profile at 1.5~GHz for NGC~891 (blue points; \citealt{2019Schmidt}) compared to the predictions from the simulation. Both the time-dependent \crest and the steady-state models produce significantly more compact profiles than observed, failing to reproduce the extended radio halo.
    \textit{Lower panel:} Vertical profile of the radio spectral index ($\alpha_\nu$ between 1.5 and 6~GHz, defined such that $L_\nu\propto\nu^{-\alpha_\nu}$). The \crest model reproduces the observed steepening of the spectral index with height, consistent with aging of advected electrons, while the steady-state model (dashed grey) yields systematically flatter spectral indices. For other simulation times, the spectral index profiles vary, while the intensity profiles remain largely unchanged (see Fig.~\ref{fig:profiles_snaps}).
    }
    \label{fig:profiles}
\end{figure}

\begin{figure}
    \centering
    \includegraphics[scale=1]{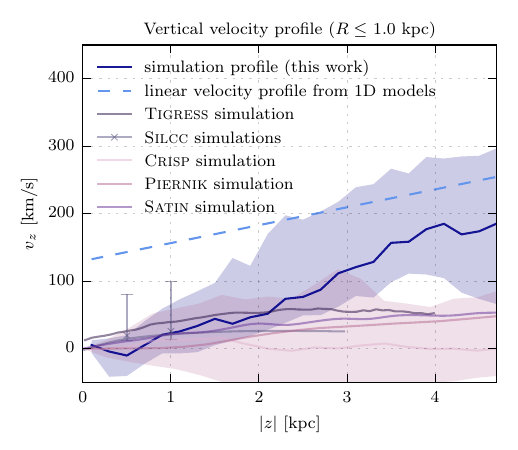}
    \caption{
    Mass-weighted vertical velocity profile within $R<1$~kpc below the galactic midplane at $t=2$~Gyr from the simulation (dark blue; for other simulation times, see Fig.~\ref{fig:profiles_snaps}). The solid lines indicate the median velocities, with shaded regions showing the 16th–84th percentile ranges. The simulated winds accelerate slowly, reaching velocities of order ${\sim}200~\mathrm{km\,s^{-1}}$ only at heights of several kiloparsecs. 
    For comparison, the dashed blue line shows the linear velocity profile commonly adopted in advection-only one-dimensional models, which assumes already large velocities (${\sim}130~\mathrm{km\,s^{-1}}$) at the midplane and further acceleration with height. 
    This is in contradiction with high-resolution simulations, e.g.\ the velocity profile of the warm gas of the \textsc{Tigress} simulations \citep{2018KimOstriker}, various \textsc{Silcc} models \citep{2016bGirichidis, 2018Girichidis}, the \textsc{Piernik} simulations \citep{2025Baldacchino-Jordan}, the \textsc{Crisp} simulation \citep{2025aThomas} as well as the \textsc{Satin} simulation \citep{2023Bieri}.
    }
    \label{fig:vertical_velocity_profile}
\end{figure}

Vertical radio continuum profiles of edge-on star-forming galaxies are often interpreted using one-dimensional (1D) CR transport models, in which CR electrons are assumed to propagate either purely by advection or by diffusion \citep{2016Heesen, 2018Heesen, 2019Schmidt}. These models typically prescribe functional forms for the vertical profiles of the magnetic field strength and, in the case of advection-dominated transport, the outflow velocity. The free parameters of the model are then adjusted to reproduce the observed radio intensity and spectral index profiles.

Using \crest, we can directly test whether such advection-only interpretations remain valid when CR electrons are evolved self-consistently in time within a full MHD galaxy simulation. In particular, we examine whether galaxies that have been classified as “advection-dominated” based on 1D modelling can be reproduced by a live electron treatment in which CR electrons are injected at SN remnants, advected with the gas, and subject to spatially and temporally varying cooling conditions. For comparison, we also consider the cell-based steady-state approach, in which CR electrons are locally tied to CR protons and inherit their diffusive transport. While this method naturally populates outflows with electrons, it assumes local equilibrium between injection and cooling even far from the electron injection sites and neglects the time-dependent evolution of the electron spectra.

A prominent example of a galaxy interpreted as advection-dominated is the above mentioned Milky Way–like edge-on galaxy NGC~891, whose vertical radio intensity and spectral index profiles were successfully fitted using a 1D advection-only model by \citet{2019Schmidt}. In Fig.~\ref{fig:profiles}, we compare the observed vertical profiles of NGC~891 with those obtained from our \crest and steady-state models at $t=2$~Gyr. We chose this time since it exhibits a particularly strong outflow towards the negative $z$-direction. For other simulation times, see App.~\ref{app:ProfilesDifferentTimes}. 
To produce the vertical profiles from the simulation, we extract a vertical strip around the center with a total width of 2.64~kpc (as done in the observations). 
We smooth the intensity maps at 1.5 and 6~GHz with a Gaussian filter with different standard deviations, corresponding to a FWHM of 0.3 and 0.5~kpc. We use the maps with smoothing of 0.5~kpc to create the spectral index profiles, mimicking the angular resolution of the observations (i.e. $12\arcsec$, corresponding to ${\sim}0.5$~kpc at the distance of NGC~891). For the intensity profile, we show in Fig.~\ref{fig:profiles} the profile with a lower smoothing with a FWHM of 0.3~kpc, since the data for the intensity profile shown in \citet{2019Schmidt} has been beam-deconvolved\footnote{Our qualitative conclusions do not depend on the exact choice of the smoothing FWHM.}. 

None of the models reproduce the observed vertical intensity profile perfectly. The steady-state model yields a somewhat more extended radio halo and agrees with the observed intensity profile up to heights of ${\sim}0.5$~kpc, while the \crest model exhibits a much stronger concentration of emission toward the disc. This behaviour is already evident in the edge-on radio maps (Fig.~\ref{fig:map_emission_alpha}) and naturally leads to the steep decline of the vertical intensity profile in the \crest case. Although the steady-state model can better reproduce the vertical extent of the emission, even this agreement depends on the evolutionary stage of the simulation and is not robust across snapshots.

The origin of these differences -- between the apparent success of the 1D advection‑only model and the much more compact profiles obtained with the live \crest modelling -- lies primarily in the assumed vertical velocity profile. In Fig.~\ref{fig:vertical_velocity_profile}, we show the linear velocity profile assumed in the 1D modelling of \citet{2019Schmidt}, which starts with a velocity of $\varv_{z,0}=130~\mathrm{km\,s^{-1}}$ already in the midplane (blue dashed line). In contrast, our simulations self-consistently develop CR-driven winds that accelerate gradually with height, reaching velocities of $\varv_z\sim200~\mathrm{km\,s^{-1}}$ only at heights of $3$–$4$~kpc. As a result, CR electrons in our simulations remain close to the disc for much longer, where magnetic field strengths, gas densities, and radiation fields are high and cooling times are short. Consequently, radio-emitting electrons cool efficiently before they can be transported to larger heights. In contrast, a fast wind launched directly from the midplane enables electrons to escape the high-loss region rapidly, allowing a substantial population of radio-emitting electrons to persist at kiloparsec scales.
To assess whether the wind acceleration profile in our simulation is model-specific, Fig.~\ref{fig:vertical_velocity_profile} also shows the vertical velocity profiles from several other simulation setups, all of which exhibit similarly slowly accelerating winds launched from the disc. We discuss these in more detail in Section~\ref{sec:DiscussionVerticalProfiles}.

The differences between the models are also reflected in the vertical spectral index profiles (shown in the lower panel of Fig.~\ref{fig:profiles}). In the steady-state model, the spectral index remains nearly constant at $\alpha_\nu \simeq 0.9$ throughout the halo, with only a slight flattening at larger heights. This behavior contrasts with the observations of NGC~891, which show a progressive steepening from $\alpha_\nu \simeq 0.7$ in the midplane to values of ${\sim}1.2$ at heights of ${\sim}2$~kpc. The \crest model qualitatively reproduces this steepening trend, with the spectral index decreasing from ${\sim}0.8$ in the disc to ${\sim}1.2$ at heights of ${\sim}1$~kpc. At larger heights, the spectral index steepens sharply, primarily because the radio intensity itself becomes negligible in the \crest model.

\begin{figure}
    \centering
    \includegraphics[scale=1]{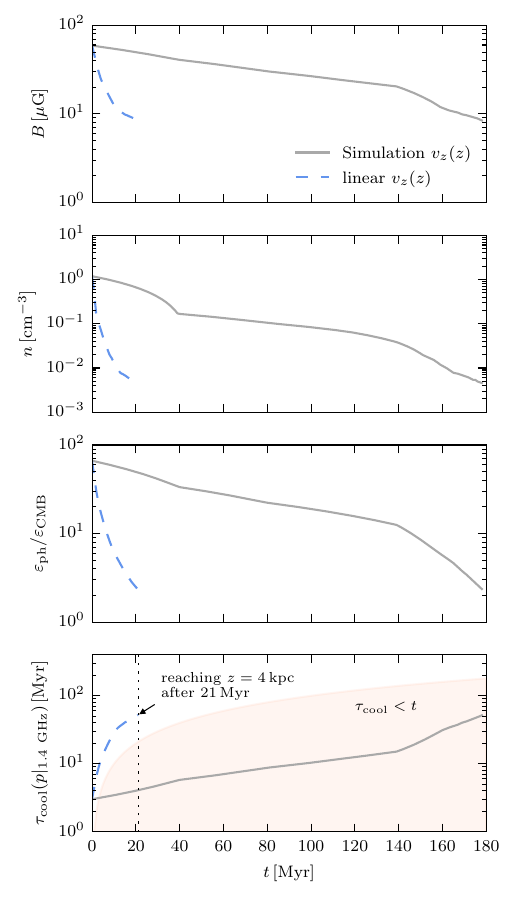}
    \caption{
    Environmental conditions experienced by a tracer particle as it is advected vertically from the disc ($z=0$) to a height of $z=4$~kpc. From top to bottom, we show the magnetic field strength, gas density, photon energy density (in units of the CMB energy density), and the cooling timescale evaluated at the characteristic electron momentum emitting GHz synchrotron radiation.  In the lower panel, the shaded region indicates where the cooling timescale is shorter than the elapsed advection time, $\tau_\mathrm{cool} <t$, corresponding to efficient radiative cooling.
    The slowly accelerating velocity profile extracted from the simulation (solid grey) keeps the tracer in dense, strongly magnetised, fast-cooling regions for an extended period of time. In contrast, adopting a linear velocity profile with high midplane velocities (dashed blue) rapidly transports the tracer out of the disc, allowing it to escape the fast-cooling regime already within ${\sim}21$~Myr.
    }
    \label{fig:vertical_time_profiles}
\end{figure}

\begin{figure*}
    \centering
    \includegraphics[scale=1]{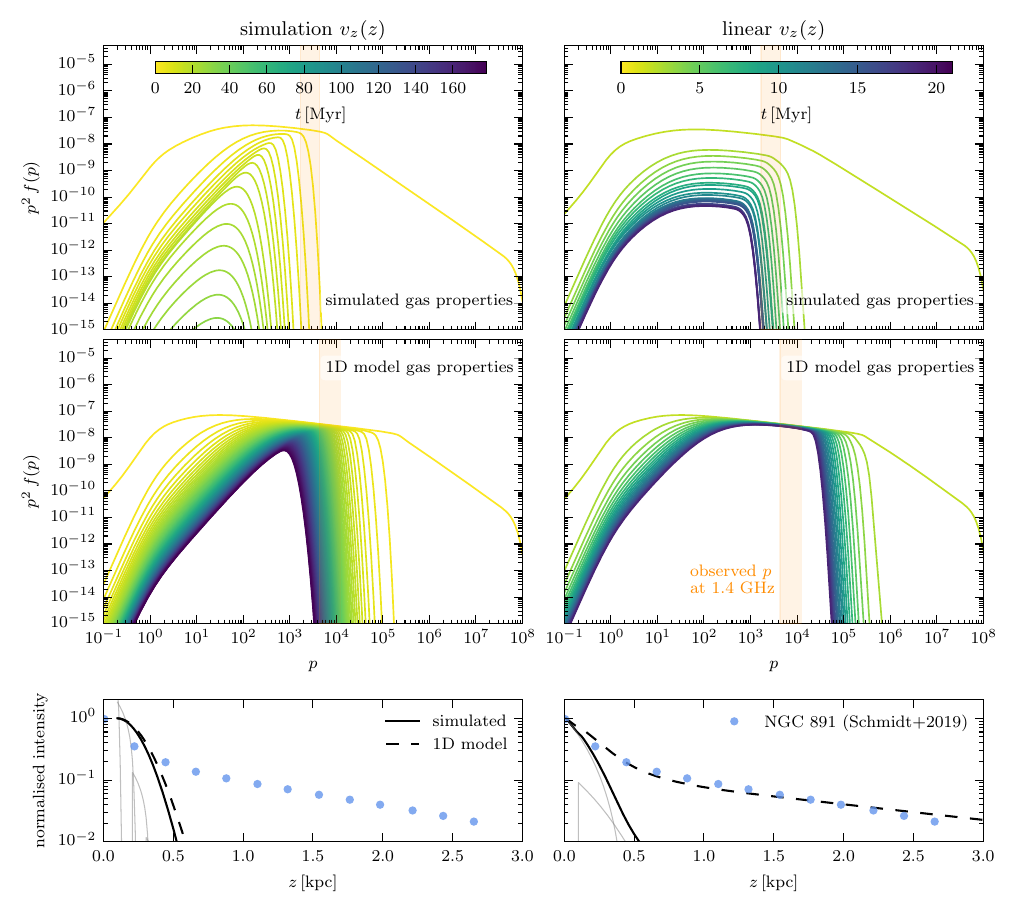}
    \caption{Time evolution of CR electron spectra along tracer trajectories assuming either a simulation-like, slowly accelerating vertical velocity profile (left-hand panels) or a linear velocity profile (right-hand panels), which starts with a large velocity in the midplane (see dashed blue line in Fig.~\ref{fig:vertical_velocity_profile}).
    The top row shows results obtained using vertical gas property profiles directly extracted from the simulation. In the middle row, the gas properties are taken from the fitted 1D model profiles \citet{2019Schmidt} (see text for details). Shaded orange regions indicate the range of electron momenta that emit into the radio band at 1.4~GHz, given the range of magnetic field strengths encountered by the tracers (see Eq.~\ref{eq:p_syn}).
    The bottom row presents the resulting normalised vertical synchrotron intensity profiles at 1.4~GHz. Grey curves show the contributions from individual tracer trajectories launched at discrete initial heights. Black curves indicate the composite intensity profiles obtained by weighting the tracer contributions according to an exponentially declining SFR-profile with height 50~pc and subsequently smoothing with a gaussian kernel with a FWHM of 0.3~kpc. Solid black lines correspond to the simulated gas properties, while dashed black lines show the results obtained with the gas properties obtained by the 1D model. The comparison illustrates that extended radio intensity profiles are reproduced only when adopting very large midplane advection speeds, as assumed in the linear velocity profiles of the 1D model.}
    \label{fig:electron_spectra_1D}
\end{figure*}

To isolate the role of the velocity profile in shaping the vertical intensity profiles, we perform a controlled tracer experiment. We extract vertical profiles of the magnetic field strength $B(z)$, gas density $n(z)$, and photon energy density $\varepsilon_\mathrm{ph}(z)$ from the simulation (within $R<1$~kpc) and convert them into time-dependent tracer trajectories by assuming two different velocity profiles: (i) a linear profile as adopted by \citet{2019Schmidt}, and (ii) the slowly accelerating profile measured directly from the simulation\footnote{For simplicity, we restrict the velocity profile in this idealised tracer experiment to positive values away from the disc midplane and impose a minimum outflow speed of $5~\mathrm{km\,s^{-1}}$. This avoids a short transient phase in the simulation during which gas at $z\simeq0.5$~kpc exhibits negative vertical velocities (see Fig.~\ref{fig:vertical_velocity_profile}).}. The resulting trajectories, together with the corresponding cooling timescales for electrons emitting at 1.4~GHz, are shown in Fig.~\ref{fig:vertical_time_profiles}.

Applying the linear velocity profile, which assumes a large midplane velocity (of $130~\mathrm{km\,s^{-1}}$), tracer particles reach a height of $z=4$~kpc within ${\sim}20$~Myr. In contrast, the simulation-based velocity profile transports electrons to the same height only after ${\sim}180$~Myr. As a consequence, electrons advected with the slower wind remain much longer in regions of strong cooling, leading to substantial energy losses before reaching the halo.

We use these trajectories as input to \crest to compute the resulting spectral evolution. The upper panels of Fig.~\ref{fig:electron_spectra_1D} show that the simulation-like velocity profile (left-hand panel) leads to strong cooling and a severe depletion of high-energy electrons at $z=4$~kpc, whereas the linear velocity profile (right-hand panel) preserves a significant population of radio-emitting electrons at this height.

Finally, we repeat this experiment using the gas property profiles consistent with those adopted in the 1D modelling of \citet{2019Schmidt}. Their fits imply substantially reduced midplane gas properties compared to our simulated galaxy, most notably a lower magnetic field strength. 
In their analysis, the vertical magnetic field profile is described by a two‑component exponential model, 
\begin{align}
    B(z) = B_1 \times \exp(-|z|/h_{B1}) + (B_0 - B_1) \times \exp(-|z|/h_{B2}),
\end{align}
with a midplane magnetic field strength of $B_0=13.2~\mathrm{\upmu G}$ and scale heights $h_{B1}=0.20$~kpc and $h_{B2}=4.4$~kpc\footnote{The value of $B_1$, which sets the relative contribution of the halo field component, is not explicitly given by \citet{2019Schmidt}. For the purpose of this experiment, we adopt  $B_1=10.5~\mathrm{\upmu G}$, which yields a reasonable match to the observed vertical intensity profile.}. 
In addition, \citet{2019Schmidt} estimate an average midplane gas density of $0.22~\mathrm{cm^{-3}}$ and a photon energy density of $\varepsilon_\mathrm{ph}=1.04\times10^{-12}\,\mathrm{erg\,cm^{-3}}\approx 2.5\times \varepsilon_\mathrm{CMB}$, both assumed to be constant with height.
Compared to our simulation, these midplane gas properties are lower than in the simulation and hence lead to longer electron cooling times near the disc. By adopting this more favourable cooling environment, we can assess whether reduced radiative losses alone are sufficient to maintain a population of radio‑emitting electrons at large heights, or whether the vertical velocity profile remains the dominant factor determining the success of advection‑only models.

As shown in the second row of panels in Fig.~\ref{fig:electron_spectra_1D}, the reduced cooling environment allows electrons to retain more energy in both cases (i.e. both for the simulation-like and the linear velocity profiles). Nevertheless, even under these conditions, the slowly accelerating velocity profile obtained from the simulation still prevents a significant population of radio-emitting electrons from reaching large heights.
This is illustrated by the orange shaded region, which indicates the range of electron momenta that emit into 1.4~GHz, given the local magnetic field strength encountered by the tracer particle (see Eq.~\ref{eq:p_syn}). A substantial overlap of this shaded region with the electron spectrum is achieved only when both the reduced gas properties from the 1D model and the assumed linear velocity profile with a high midplane velocity are adopted.

The third row in Fig.~\ref{fig:electron_spectra_1D} shows the resulting normalized vertical intensity profiles obtained from the tracer experiments. The grey curves show the contributions from individual tracer trajectories launched at discrete initial heights above the disc midplane ($z_0=0.1$-$0.5$~kpc), while the black curve represents the composite emissivity profile. The latter is constructed by weighting and smoothing the individual tracer contributions so as to approximate a continuous vertical distribution of CR electron sources. 
Specifically, the profiles are weighted according to an exponentially declining SFR distribution in the vertical direction, i.e. $\exp^{-|z|/h_\mathrm{SFR}}$, with a scale height $h_\mathrm{SFR}=0.05$~kpc\footnote{Our qualitative conclusions are insensitive to the exact choice of this value.}.
The dashed curve illustrates the case in which the gas properties are reduced to those adopted in the 1D modelling of \citet{2019Schmidt}. 

Even under these more favourable cooling conditions, the vertical intensity profile for the simulation-like, slowly accelerating velocity field drops sharply beyond $z \gtrsim 0.5$~kpc. Adopting a linear velocity profile leads to a somewhat more extended intensity distribution. However, an extended profile consistent with the observations is obtained only when both the reduced gas properties and the linear velocity profile assumed in the 1D model are adopted

This comparison demonstrates that the assumed vertical velocity profile is the dominant factor determining whether advection-only models can reproduce extended radio haloes. In Milky Way–mass galaxies with self-consistently driven, slowly accelerating winds, advection alone is therefore insufficient to explain the observed vertical extent of radio emission, even when adopting favourable cooling conditions.


\section{Does the radio spectrum trace the galaxy-wide CR electron spectrum?}\label{Sec:GlobalRadioGlobalElectronSpectra}

A common assumption in the interpretation of radio observations is that the integrated observed synchrotron spectrum directly reflects the underlying, galaxy-wide volume-weighted CR electron spectrum. In this section, we test this assumption by comparing CR electron spectra and the resulting radio synchrotron spectra obtained with different modelling approaches: a simple one-zone steady-state model, a cell-based steady-state model implemented with \crayon, and the live CR electron modelling with \crest.

\begin{figure*}
    \centering
    \includegraphics[]{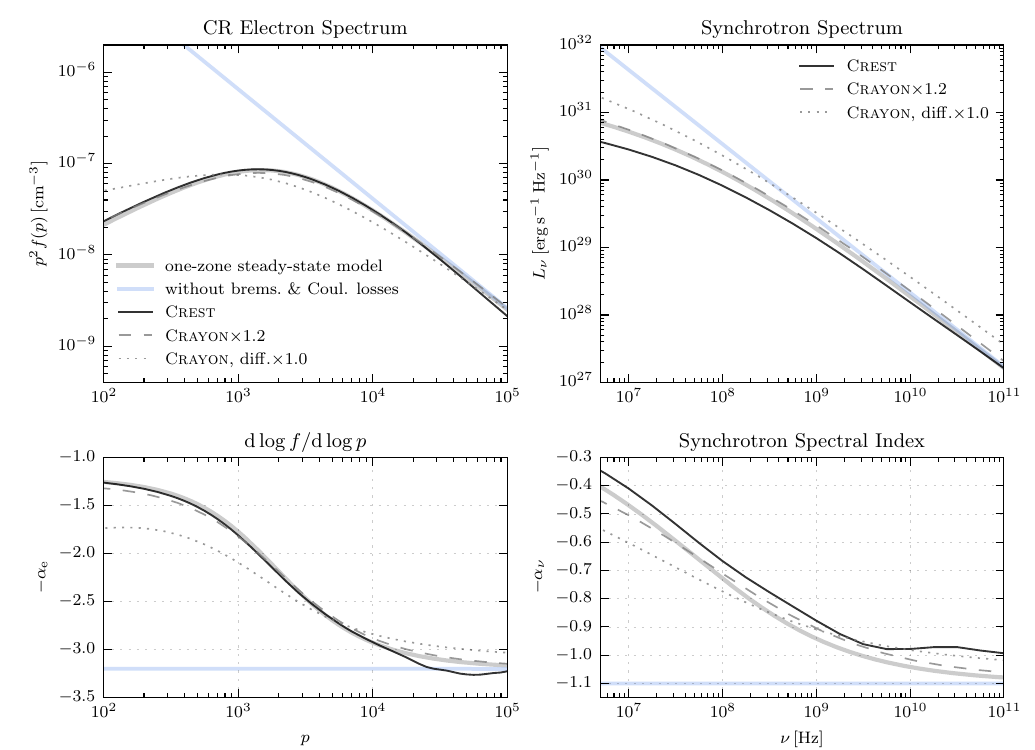}
    \caption{
    Galaxy-integrated CR electron spectra (left panels) and the corresponding radio synchrotron spectra (right panels) at $t=1$~Gyr for different modelling approaches, comparing time-dependent modelling with \crest to steady-state modelling with \crayon, shown both with diffusion (dotted lines) and without diffusion (dashed lines). The latter are rescaled by a factor of 1.2 to facilitate a direct comparison of spectral shapes. The lower panels show the corresponding spectral slopes. 
    The CR electron spectra are nearly identical between the one-zone steady-state model (grey), the cell-based steady-state model without diffusion (dashed), and the live electron modelling with \crest (black) over the momentum range relevant for radio emission. 
    Despite this, the resulting synchrotron spectrum is systematically flatter in the \crest model. 
    Neglecting bremsstrahlung and Coulomb losses (blue) yields a steep electron spectrum with slope $\simeq -3.1$, which translates into a uniformly steep synchrotron spectrum with $\alpha_\nu \simeq 1.1$ across all frequencies, as expected in the fully cooled synchrotron/IC limit.
    }
    \label{fig:electron_synchrotron_spectra}
\end{figure*}

Figure~\ref{fig:electron_synchrotron_spectra} shows the galaxy-integrated, volume-weighted CR electron spectra and the corresponding face-on synchrotron spectra at $t=1$~Gyr. For the cell-based steady-state approach, we show two variants: one including spatial diffusion in the steady-state equation and one neglecting diffusion. The latter provides the most direct comparison to the one-zone steady-state model and the \crest results, both of which do not include explicit electron diffusion.

At electron momenta relevant for GHz radio emission in $\upmu$G magnetic fields, $p \sim 10^{3}$–$10^{4}$, the CR electron spectra predicted by \crest, the one-zone steady-state model, and the cell-based steady-state model without diffusion are nearly indistinguishable. Only the cell-based steady-state model including diffusion shows a noticeable deviation in this momentum range. At higher momenta, $p \gtrsim 10^{4}$, the \crest spectrum steepens relative to the steady-state models, reflecting the explicit time-dependent treatment of electron cooling along tracer trajectories \citep{2025Werhahn_CREST}.

\begin{figure*}
    \centering
    \includegraphics[]{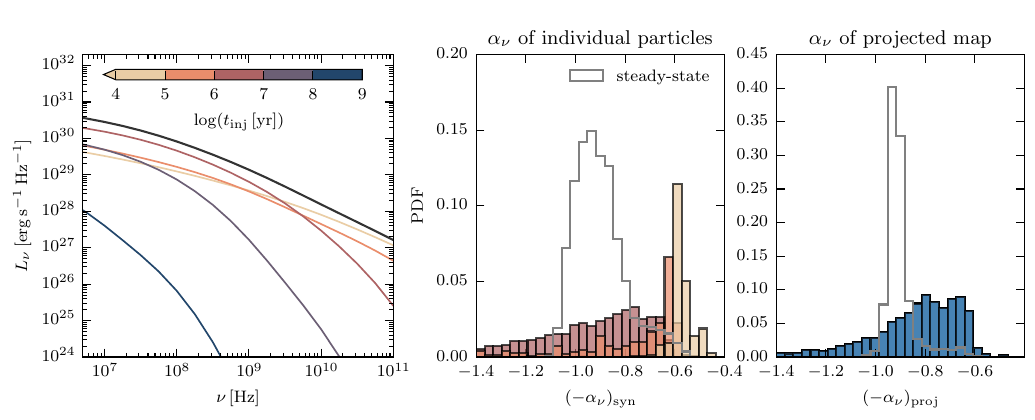} 
    \caption{
    Age-resolved contributions to the radio synchrotron emission in the \crest model at $t=1$~Gyr.
    \textit{Left:} Decomposition of the galaxy-integrated synchrotron spectrum (black) into contributions from CR electrons binned by the time since their last injection event, $t_\mathrm{inj}$ (colour-coded). At frequencies $\lesssim 1$~GHz, the emission is dominated by electrons injected $10^{6}$–$10^{7}$~yr ago, while at higher frequencies the youngest electron population ($t_\mathrm{inj}<10^{5}$~yr) dominates.
    \textit{Middle:} Luminosity-weighted distribution of face-on synchrotron spectral indices between 140~MHz and 1.4~GHz for individual \crest tracer particles (colour-coded by age) and for cells in the steady-state model (grey). The steady-state distribution peaks around $\alpha_\nu \simeq 0.9$, whereas young electrons in \crest exhibit significantly flatter spectra, with the youngest tracers clustering around $\alpha_\nu \sim 0.6$, reflecting their injection slope.
    \textit{Right:} Distribution of spectral indices derived from projected face-on intensity maps at 140~MHz and 1.4~GHz. The broader distribution in \crest demonstrates that the age-dependent spectral variations are preserved in projection, while the steady-state model yields a much narrower distribution with on average steeper indices.}
    \label{fig:radio_age_spectral_indices}
\end{figure*}

Despite these similarities in the CR electron spectra at radio-emitting momenta, the resulting synchrotron spectra differ systematically. In particular, we find two noticeable differences: 
First, the total synchrotron spectrum predicted by \crest is noticeably flatter than those obtained with the steady-state models (see lower right-hand panel in Fig.~\ref{fig:electron_synchrotron_spectra}), and second, the radio spectral index distribution is much broader in the \crest model, reflecting the age dependence of the emitting electron populations (see Fig.~\ref{fig:radio_age_spectral_indices}).

Regarding the first point, we find that, for example, at $\nu=1$~GHz, the \crest spectrum exhibits a slope of $\alpha_\nu \approx 0.88$, compared to $\alpha_\nu \approx 0.91$ and $\approx 0.94$ for the cell-based and one-zone steady-state models, respectively (see Fig.~\ref{fig:electron_synchrotron_spectra}). Interpreting these radio spectra in isolation would therefore lead to the inference of different underlying CR electron spectra, even though the actual electron spectra in the relevant momentum range ($p \sim 10^{3}$–$10^{4}$) are nearly identical.

Furthermore, Fig.~\ref{fig:electron_synchrotron_spectra} shows that all models exhibit a pronounced flattening of the synchrotron spectrum at frequencies below $\sim 100$~MHz. As illustrated by the blue curve, this behaviour is driven entirely by the inclusion of bremsstrahlung and Coulomb losses. When these processes are neglected, the one-zone steady-state model yields the canonical fully cooled electron spectral slope of $3.2$, corresponding to a constant synchrotron slope of $\alpha_\nu \approx 1.1$ across all frequencies. We return to the role of these loss processes in more detail in Section~\ref{sec:BremsstrahlungCoulombLosses}.

\begin{figure*}
    \centering
    \includegraphics[]{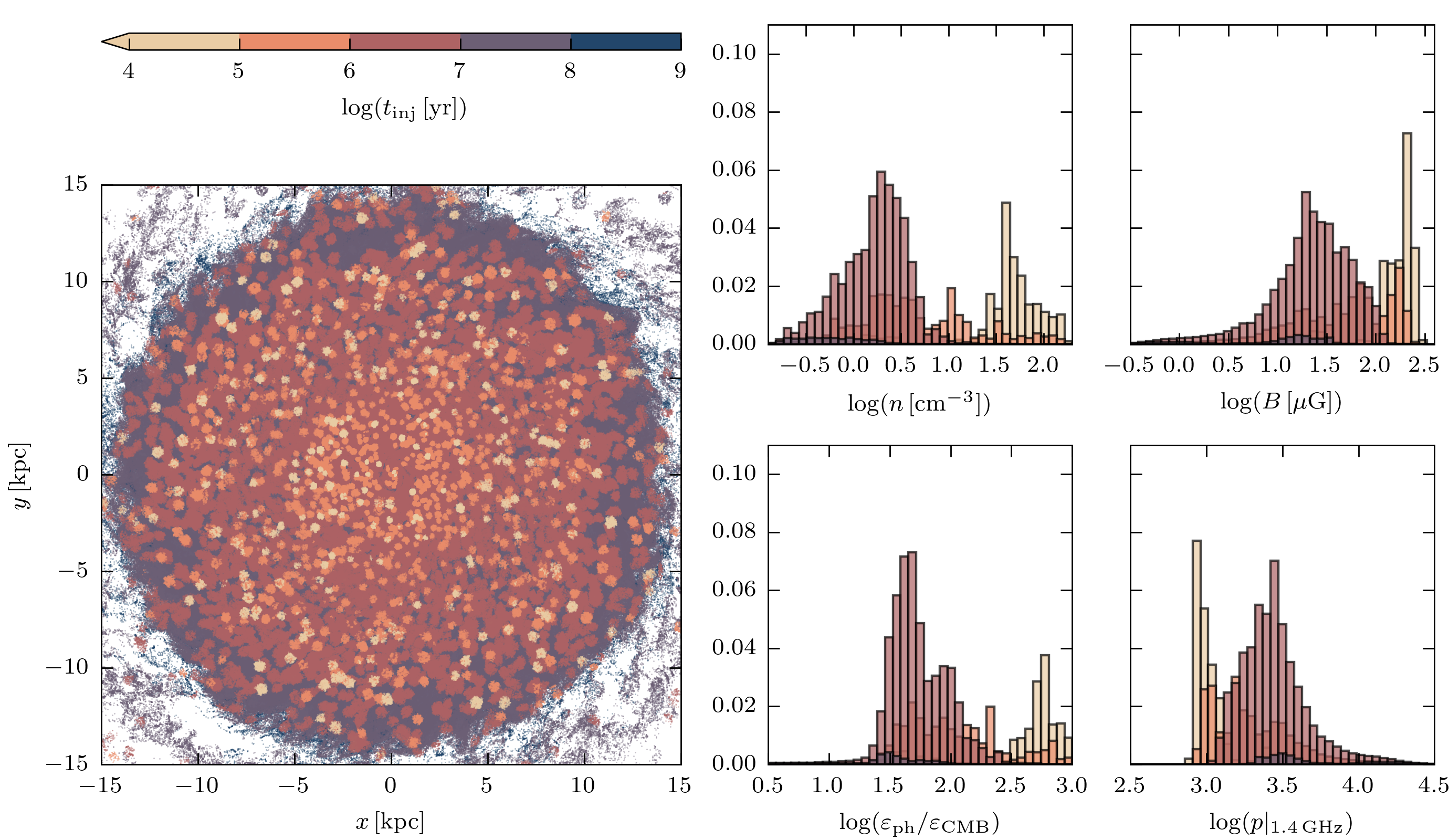}
    \caption{
    Spatial distribution and local environments of CR electron tracer particles in the \crest model at $t=1$~Gyr.
    \textit{Left:} Face-on positions of tracer particles within the disc ($|z|<1$~kpc), colour-coded by the time since their last injection event $t_\mathrm{inj}$ (older tracers are plotted first, with younger tracers overlaid to highlight recent injection sites). Young electrons remain concentrated near their SN injection sites, while older populations are more spatially extended.
    \textit{Right:} Synchrotron-luminosity–weighted distributions of gas properties sampled by the tracer particles, again colour-coded by $t_\mathrm{inj}$. 
    Young electrons preferentially reside in dense, highly magnetised regions with large photon energy densities, causing 1.4~GHz emission to originate from lower electron momenta ($p \sim 10^{3}$, see Eq.~\ref{eq:p_syn}). Older electrons sample lower-density, weaker magnetic-field environments, requiring higher momenta ($p \sim 10^{3.5}$) to emit at the same frequency. This variation in local conditions naturally produces a broad range of radio spectral indices across the electron population.
    }
    \label{fig:map_position_histograms}
\end{figure*}

To understand the origin of the flatter synchrotron spectrum in the \crest model, we next examine the contributions from CR electron populations of different ages. Figure~\ref{fig:radio_age_spectral_indices} decomposes the \crest synchrotron spectrum into contributions from tracer particles binned by the time since their last energy injection event, $t_\mathrm{inj}$. At frequencies below ${\sim} 1$~GHz, the emission is dominated by electrons with injection events $10^{6}$–$10^{7}$~yr ago, whereas at higher frequencies the emission is increasingly dominated by very young electrons with $t_\mathrm{inj} \lesssim 10^{5}$~yr. Because these youngest electrons have had less time to cool, they naturally produce flatter synchrotron spectra at high frequencies. 

This behaviour is explicitly reflected in the distribution of spectral indices shown in the middle panel of Fig.~\ref{fig:radio_age_spectral_indices}. The luminosity-weighted spectral indices of individual \crest tracer particles, computed between 140~MHz and 1.4~GHz, span a broad range and correlate strongly with electron age. The youngest electrons cluster around $\alpha_\nu \sim 0.6$, corresponding to the injected electron spectral slope of 2.2. In some cases, even flatter spectra with $\alpha_\nu \lesssim 0.4$ occur, originating from regions with very strong magnetic fields where low-energy electrons affected by Coulomb losses contribute to the radio emission. Older electron populations exhibit progressively steeper spectral indices and a broader distribution. In contrast, the cell-based steady-state model yields a narrow distribution centred around $\alpha_\nu \sim 0.9$, reflecting the assumption of local equilibrium between injection and cooling.

The right-hand panel of Fig.~\ref{fig:radio_age_spectral_indices} shows that this behaviour persists in projection. Spectral indices derived from face-on intensity maps between 140~MHz and 1.4~GHz exhibit a much broader distribution in \crest than in the steady-state model, with a substantial fraction of pixels showing flatter spectra up to $\alpha_\nu \sim 0.6$.

Fig.~\ref{fig:map_position_histograms} further illustrates how these age-dependent spectral signatures arise from the spatial distribution of CR electrons. The left panel shows the positions of tracer particles in the disc, colour-coded by $t_\mathrm{inj}$. Young electrons remain concentrated near their SN injection sites, whereas older electrons are more spatially extended. The right-hand panels show synchrotron-luminosity-weighted distributions of local gas properties sampled by the tracers. Young electrons preferentially reside in dense, highly magnetised regions with large photon energy densities, where electrons of relatively low momenta, $p \sim 10^{3}$, emit at 1.4~GHz. In contrast, older electrons inhabit regions with lower densities and magnetic field strengths, requiring higher momenta, $p \sim 10^{3.5}$, to emit at the same frequency. This variation in local conditions naturally leads to a broad range of synchrotron spectral indices for different electron populations, as seen in Fig.~\ref{fig:radio_age_spectral_indices}.

Taken together, these results demonstrate that the galaxy-integrated synchrotron spectrum does not provide a one-to-one tracer of the underlying CR electron spectrum. Instead, radio emission is biased toward young, recently injected electron populations and reflects a convolution of electron age, local cooling conditions, and transport history. As a result, radio spectra can be shaped by local variations in the underlying CR electron population, even when the volume-averaged, galaxy-wide electron spectra at radio-emitting momenta appear similar.

\section{Are bremsstrahlung / Coulomb losses negligible for radio-emitting electrons?}\label{sec:BremsstrahlungCoulombLosses}

\begin{figure*}
    \centering
    \includegraphics[]{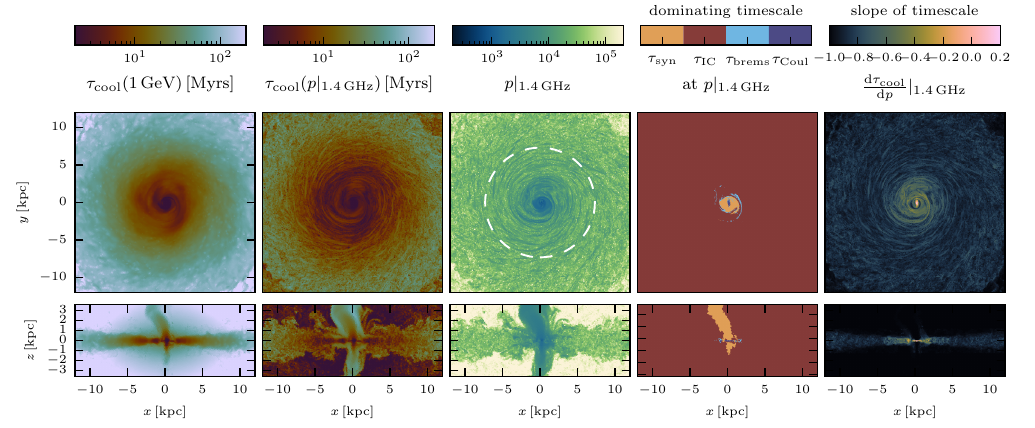}
    \caption{Spatially resolved electron cooling properties at $t=1$~Gyr, derived from slices of the relevant gas properties.
    \textit{First panel:} Cooling timescale of 1~GeV electrons (including synchrotron, IC, bremsstrahlung and Coulomb losses).
    \textit{Second panel:} Cooling timescale evaluated at the electron momentum that emits synchrotron radiation at 1.4~GHz, given the local magnetic field strength.
    \textit{Third panel:} Corresponding characteristic electron momentum for 1.4~GHz emission. The dashed white circle encloses 99\% of the total 1.4~GHz synchrotron luminosity.
    \textit{Fourth panel:} Locally dominant cooling process at the 1.4~GHz-emitting momentum (see Eq.~\ref{eq:p_syn}). Most of the disc is dominated by IC or synchrotron losses.
    \textit{Fifth panel:} Local slope of the cooling timescale, $\mathrm{d}\tau_\mathrm{cool}/\mathrm{d}p$, evaluated at the same momentum. While synchrotron and IC losses alone would yield a slope of $-1$ (black), large regions of the disc, especially the central regions dominating the radio emission, exhibit significantly flatter slopes, indicating a non-negligible influence of bremsstrahlung and Coulomb losses on the shape of the electron spectrum.}
    \label{fig:map_timescales}
\end{figure*}

It is often assumed that radio emission at GHz frequencies predominantly traces CR electrons with GeV energies, for which synchrotron and IC losses typically dominate over bremsstrahlung and Coulomb losses. While this assumption is broadly correct in $\upmu$G magnetic fields, it does not necessarily imply that bremsstrahlung and Coulomb losses are irrelevant for shaping the radio-emitting electron population. In this section, we demonstrate that these processes can significantly influence the electron spectrum -- and hence the radio synchrotron spectrum -- even when they do not dominate the total energy loss rate.

Figure~\ref{fig:map_timescales} illustrates this point by showing spatially resolved cooling timescales from slices of the simulation at $t=1$~Gyr. The left-hand panel shows the cooling timescale of 1~GeV electrons across the galactic disc. However, because the magnetic field strength varies substantially throughout the galaxy, electrons emitting at a fixed observing frequency do not have a uniform characteristic momentum.

This is shown explicitly in the second and third panels of Fig.~\ref{fig:map_timescales}, which display the cooling timescale and the corresponding electron momentum for synchrotron emission at 1.4~GHz, given the local magnetic field strength. The cooling timescales at the relevant emitting momenta are shortest (of order a few Myrs) in the central, strongly magnetised regions of the disc with large photon energy densities and increase with galactocentric radius, reflecting the decline in both magnetic field strength and radiation energy density. Within the region that contributes 99\% of the total 1.4~GHz emission (indicated by the dashed white circle), the characteristic momenta range from $p\sim10^{3}$ in the strongly magnetised central regions to $p\sim10^{4}$ in the outer disc.

The fourth panel shows which cooling process dominates locally at the momentum corresponding to 1.4~GHz emission.
Most of the disc is IC-dominated (shown in red), while synchrotron losses dominate in the central regions and in highly magnetised outflows (shown in orange). Bremsstrahlung and Coulomb losses dominate only in small, localized patches and are therefore often considered negligible for radio-emitting electrons.

However, the final panel demonstrates why this conclusion is incomplete. While synchrotron and IC losses both imply cooling timescales scaling as $\tau_\mathrm{cool} \propto p^{-1}$, bremsstrahlung and Coulomb losses have flatter momentum dependencies. The map of the local slope $\mathrm{d}\tau_\mathrm{cool}/\mathrm{d}p$ at the 1.4~GHz-emitting momentum shows that much of the disc exhibits slopes significantly flatter than $-1$, particularly in the central regions where most of the radio emission originates.

As a result, even when bremsstrahlung and Coulomb losses are sub-dominant in terms of absolute cooling timescales, they still modify the effective cooling behaviour of the electron population. This leads to a flattening of the cooled electron spectrum at low and intermediate momenta ($p \lesssim 10^{5}$), which in turn produces the pronounced flattening of the synchrotron spectrum below ${\sim}100$~GHz seen in Fig.~\ref{fig:electron_synchrotron_spectra}. 
At lower observing frequencies, such as 144~MHz, which probe lower-energy electrons, the influence of bremsstrahlung and Coulomb losses on the radio spectrum is expected to be even more pronounced.
We therefore conclude that bremsstrahlung and Coulomb losses cannot be neglected when interpreting detailed shapes of radio spectra, particularly in dense and strongly magnetised regions of star-forming galaxies, which dominate the total radio emission.

\section{Discussion}

\subsection{Effects of free-free absorption and emission}
\label{sec:ff-emission}

\begin{figure}
    \centering
    \includegraphics[]{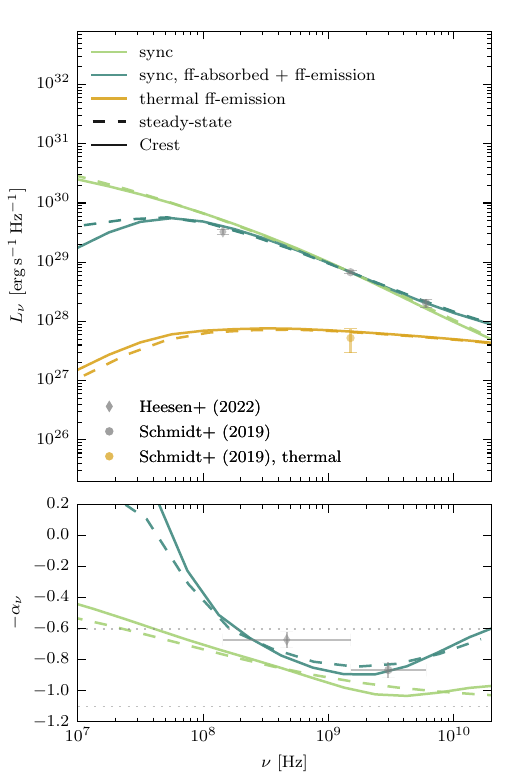}
    \caption{
    Normalised radio continuum spectra of NGC~891 viewed edge-on, comparing the \crest (solid lines) and steady-state (dashed lines) models at $t=1$~Gyr. The spectra are normalised to the observed total flux density at 1.5~GHz \citep{2019Schmidt} by factors of 0.8 (steady-state) and 1.5 (\crest). 
    \textit{Top panel:} Total luminosity spectrum as a function of frequency. The unabsorbed synchrotron emission (green) is shown alongside the spectrum including free-free absorption and thermal free-free emission (blue). Observed data points are from \citet{2019Schmidt} at 1.5 and 6~GHz and \citet{2022Heesen} at 144~MHz, with total (grey) and thermal (gold) contributions shown separately. 
    \textit{Bottom panel:} Spectral index $\alpha_\nu$ (defined as $L_\nu \propto \nu^{-\alpha_\nu}$) as a function of frequency. The grey points indicate spectral indices derived from the observed flux densities at 144~MHz, 1.5~GHz, and 6~GHz shown in the upper panel. The dotted horizontal lines indicate the theoretical spectral indices of $\alpha_\nu =(\alpha_\mathrm{e} - 1)/2 =0.6$ and $1.1$ for the injected ($\alpha_\mathrm{e} = 2.2$) and fully cooled ($\alpha_\mathrm{e} = 3.2$) electron spectra, respectively. 
    While free-free absorption strongly suppresses the total emission only at frequencies $\lesssim 500$~MHz, and thermal free-free emission becomes significant only at $\gtrsim 5$~GHz, the spectral index is affected across the entire frequency range in both models. The effect is smallest around $1$~GHz, where the spectrum most closely reflects the underlying synchrotron emission.}
    \label{fig:spectra_ff_abs_ff_em}
\end{figure}

In this work, we have analysed radio synchrotron diagnostics commonly applied to observations of star-forming galaxies, but we have so far neglected the effects of free-free absorption and emission. \citet{2021WerhahnIII} found, in a similar isolated galaxy simulation setup, that the radio spectra of nearby star-forming galaxies such as NGC~253 and M82 can be well reproduced by accounting for free-free absorption, which flattens the spectrum at frequencies $\lesssim 1$~GHz, and free-free emission, which contributes significantly above $\gtrsim 10$~GHz. While synchrotron emission was found to dominate around $\sim$~GHz frequencies, the spectral shape was already affected at and below those frequencies.

Modelling these effects accurately would require a multi-phase ISM, which is beyond the scope of our current setup. Instead, we use the sub-grid pressurised ISM model of \citet{2003SpringelHernquist}, which provides an effective free-electron fraction, and compute the corresponding free-free emission and absorption assuming a warm ionised medium temperature of $T=8000$~K \citep[see Appendix~A2 of][for the relevant equations]{2021WerhahnIII}.

Under these assumptions, we present a representative snapshot at $t=1$~Gyr in Fig.~\ref{fig:spectra_ff_abs_ff_em}. To facilitate comparison of spectral shapes, all spectra are renormalised to the observed total flux density at 1.5~GHz. The resulting normalisation factors for the total \crest and steady-state spectra (including free-free absorption and emission) are 1.5 and 0.8, respectively. To match the thermal free-free emission estimate of \citet{2019Schmidt}, i.e.\ the golden data point in Fig.~\ref{fig:spectra_ff_abs_ff_em}, we apply a constant rescaling factor $\xi_\mathrm{e}$ to the free-electron density provided by the sub-grid model, obtaining $\xi_\mathrm{e} = 0.4$ and 0.3 for the \crest and steady-state models, respectively. This brings the modelled thermal contribution into agreement with the observed value and simultaneously improves the agreement of the total spectral shape with the data points at 144~MHz and 6~GHz.

The lower panel of Fig.~\ref{fig:spectra_ff_abs_ff_em} shows the radio spectral index as a function of frequency. In both models, the unabsorbed synchrotron spectrum steepens from $\alpha_\nu\approx 0.5$ at 10~MHz to $\alpha_\nu \approx 1$ at 10~GHz, remaining flatter than the theoretical value of $\alpha = 1.1$ expected for a fully cooled electron population with spectral index $\alpha_\mathrm{e} = 3.2$. As discussed in Section~\ref{Sec:GlobalRadioGlobalElectronSpectra}, the \crest spectrum is flatter than the steady-state spectrum at frequencies $\lesssim 100$~MHz, and exhibits small temporal spectral variations at higher frequencies (see Fig.~\ref{fig:radio_spectra}). The substantial spectral flattening below ${\sim} 10$~GHz in both models can be attributed to bremsstrahlung and Coulomb losses, as demonstrated in Section~\ref{sec:BremsstrahlungCoulombLosses}.
The inclusion of free-free absorption flattens the spectra further at low frequencies in both models, bringing them into better agreement with the observed data. Crucially, we find that the spectral modification due to free-free absorption at low frequencies is substantially larger than the difference in spectral slope between the two modelling approaches, underlining the importance of accounting for thermal absorption effects when interpreting low-frequency radio observations of star-forming galaxies.

\subsection{Implications of the vertical radio intensity profiles}\label{sec:DiscussionVerticalProfiles}

The mismatch between the vertical radio intensity profiles predicted by \crest and those inferred from observations has important implications for our understanding of CR electron transport in star-forming galaxies. The \crest framework provides a highly accurate description of the time-dependent evolution of CR electron spectra along Lagrangian tracer particles and therefore captures, by construction, the advection of electrons in a full MHD galaxy simulation with self-consistently amplified magnetic fields and CR-driven winds.
At the same time, the \crest results successfully reproduce several key observational constraints. In particular, the global radio spectrum and total radio luminosity are in good agreement with observations, and the characteristic steepening of the radio spectral index with height naturally emerges from electron aging as particles are transported away from their injection sites. This provides strong support to the physical realism of the underlying electron cooling and emission modelling.

However, despite this success, the advection-only \crest model fails to reproduce the observed extended vertical radio intensity profile of the observed MW-like galaxy NGC~891 (as discussed in Section~\ref{sec:VerticalProfilesAdvection}). This discrepancy points to two broad classes of possible explanations.

A first possibility is that the winds launched in our isolated galaxy simulation accelerate too slowly compared to real galactic winds. If winds in nature reach high velocities much closer to the midplane, advection alone could in principle transport radio-emitting electrons to larger heights before they cool (as we demonstrated in Fig.~\ref{fig:electron_spectra_1D}). 
However, there are physical arguments that disfavour winds reaching very large bulk velocities already at very small heights above the disc in Milky Way–like galaxies. Feedback‑driven outflows are powered by energy and momentum injection that is vertically distributed over the scale height of star formation and SNe -- typically of order a few hundred parsecs. While individual SN remnants drive highly supersonic shocks, the large‑scale, volume‑filling wind is set by the resulting pressure gradients and therefore accelerates gradually over this vertical extent. As a consequence, the bulk outflow velocity near the midplane is expected to be subsonic or transonic, with higher velocities reached only at larger heights. Pressure‑driven wind models indeed predict such gradual acceleration, although the launch velocity close to the disc depends sensitively on the assumed boundary conditions and feedback prescriptions \citep[e.g.\ ][]{1991Breitschwerdt, 2008Everett, 2016Recchia}.
We note that more rapid acceleration close to the disc may occur in extreme starburst systems \citep[such as M82, see e.g.\ ][]{2015Leroy}, which differ substantially from the Milky Way-like galaxies considered here.
Furthermore, we emphasize that the velocity profiles discussed here are mass-weighted and therefore dominated by the cold and warm gas phases that carry most of the mass in our simulation. Within the advection-dominated framework considered here, CR electrons are assumed to be transported with this gas and thus inherit its kinematics. This does not preclude that, in reality, CR electrons could also be transported by hotter, low-density gas phases that may reach higher velocities at smaller heights, or that additional transport processes allow CR electrons to decouple from the bulk gas flow, as discussed below.

Slowly accelerating winds are also a common outcome of a wide range of other simulation setups, spanning both global galaxy simulations and stratified-box setups (see Fig.~\ref{fig:vertical_velocity_profile}). This behaviour is seen in other global galaxy simulations that account for a multiphase ISM and SN-driven feedback \citep{2023Bieri}, or that employ a two-moment CR transport scheme combined with a pressurised ISM prescription \citep{2023Thomas} or accounting for a multiphase ISM \citep{2025aThomas}. Similar behaviour is found in very high-resolution stratified-box simulations from the \textsc{Silcc} project \citep{2016bGirichidis, 2018Girichidis}, the \textsc{Tigress} simulations \citep{2018KimOstriker}, as well as the stratified-box simulations performed with the \textsc{Piernik} code \citep{2025Baldacchino-Jordan}. 
While it remains possible that current simulations collectively miss some key physics relevant for wind launching and acceleration, the consistency of this result across methodologies suggests that unrealistically slow winds are unlikely to be the sole explanation.

A second, and in our view more plausible, interpretation is that advection alone is insufficient to reproduce the observed vertical extent of radio haloes in star-forming galaxies. One-dimensional models that successfully match the observed intensity profiles using advection-only transport typically require very high wind velocities already near the midplane. If such velocity profiles are unrealistic, this implies that additional transport or acceleration mechanisms must contribute. Possible candidates include spatial diffusion, CR streaming, or in-situ re-acceleration processes.

Additional transport processes are particularly compelling at larger heights above and below the disc, where electron cooling timescales become long. In these regions, CR transport relative to the gas may become increasingly important, leading to more extended radio-emitting electron populations. In practice, such transport arises from pitch‑angle scattering of CRs off self‑generated Alfvén waves or pre‑existing MHD turbulence, which results in a combination of streaming and diffusion in the wave frame. While the effective transport coefficients remain poorly constrained, it is plausible that this CR transport relative to the gas could significantly extend the radio-emitting electron population to larger heights. If this interpretation is correct, the vertical extent and spectral index profiles of radio haloes offer a powerful observational tool for constraining CR transport physics in galactic haloes.

This is supported by the finding that in our steady-state modelling with \crayon, the intensity profiles at larger heights are much closer to the observations. In this approach, the CR electron population is tied to the CR protons in each computational cell in post-processing, which are not only advected with the gas but also allowed to diffuse along magnetic field lines. As shown in \citet{2024ChiuSandy}, combining this approach with a two-moment CR transport scheme and a multiphase ISM model leads to a good agreement with observed vertical radio intensity profiles and even recovers the observed X-shaped morphology seen in polarized emission of NGC~4217 \citep{2020Stein}. This corroborates the conclusion that some other transport mechanism relative to the gas is required. Furthermore, this implies that edge-on radio observations of star-forming galaxies are the ideal test case to study and constrain properties of CR transport.

Another plausible explanation that could either explain or contribute to the extended radio emission is re-acceleration. As demonstrated in \citet{2025Werhahn_CREST}, the detailed \crest modelling predicts substantial populations of old, cooled electron populations in the outflows. These electrons could potentially be re-accelerated by shocks associated with galactic winds (Fermi-I acceleration), by turbulence in the halo (Fermi-II acceleration), or magnetic reconnection. We note, however, that re‑acceleration is not strictly required in all models, as extended radio haloes can also be reproduced without it \citep[e.g.][]{2024ChiuSandy}.
Whether such processes can re-energise electrons sufficiently to produce detectable radio or even $\gamma$-ray emission remains an open question and will be explored in future work.

Finally, the role of the galactic environment should be considered. The simulations analysed here model isolated galaxies, whereas real galaxies reside in a cosmological context. A more realistic circumgalactic medium may significantly alter the structure, collimation, and acceleration of galactic outflows, with corresponding consequences for CR transport and radio emission. Extending this analysis to cosmological simulations will therefore be an important next step.

\section{Conclusion}\label{sec:conclusions}
Radio continuum observations are a powerful and widely used diagnostic of CR electron populations and transport in star-forming galaxies. Their interpretation, however, rests on several simplifying assumptions whose validity has remained difficult to assess. In this work, we revisited three such assumptions by modelling radio synchrotron emission from a Milky Way–mass galaxy using time-dependent, spectrally resolved CR electron evolution with the \crest framework, and by comparing the resulting radio observables to commonly employed steady-state models.
In particular, \crest follows CR electrons injected at SN sites and subsequently advected with the gas while evolving their spectra in time, whereas the steady‑state post‑processing with \crayon ties CR electrons locally to the CR proton population and assumes injection–cooling equilibrium.

First, we tested whether vertical radio intensity profiles of edge-on galaxies can be accounted for solely by advective CR transport. While the live electron modelling with \crest naturally reproduces the observed steepening of radio spectral indices with height as a consequence of electron aging, it fails to produce the extended radio haloes observed in galaxies such as NGC~891 (see Fig.~\ref{fig:profiles}). The underlying reason is that self-consistently driven galactic winds in our simulations accelerate only gradually (Fig.~\ref{fig:vertical_velocity_profile}), keeping CR electrons in dense, strongly magnetised environments for too long, where radiative losses efficiently deplete the radio-emitting population (Figs.~\ref{fig:vertical_time_profiles} and \ref{fig:electron_spectra_1D}). Reproducing extended radio haloes with CR advection alone requires wind velocity profiles with unrealistically high midplane speeds. This strongly suggests that additional transport or acceleration processes such as diffusion, streaming, or re-acceleration are required to explain observed vertical radio profiles in star-forming galaxies.

Second, we examined whether radio spectra directly trace the underlying, galaxy-wide CR electron spectrum. Although the different modelling approaches predict nearly identical volume-weighted CR electron spectra at momenta relevant for GHz radio emission (see Fig.~\ref{fig:electron_synchrotron_spectra}), the resulting synchrotron spectra differ systematically. In particular, the \crest model produces on average flatter and more variable radio spectra with a broader distribution of spectral indices than steady-state models (Figs.~\ref{fig:radio_spectra} and \ref{fig:radio_age_spectral_indices}). We showed that this discrepancy arises because radio emission is biased toward young, recently injected electron populations residing in dense, strongly magnetised regions, where local cooling conditions and electron ages strongly shape the observed spectrum (Fig.~\ref{fig:map_position_histograms}). As a result, the galaxy-integrated radio spectrum represents a convolution of spatially-varying electron age, transport history, and local environment, rather than being a direct tracer of the global CR electron spectrum.

Third, we demonstrated that bremsstrahlung and Coulomb losses significantly influence radio spectra, even when synchrotron and IC losses dominate the total cooling rate at the relevant electron energies (Figs.~\ref{fig:electron_synchrotron_spectra} and \ref{fig:map_timescales}). By modifying the momentum dependence of the effective cooling timescale, these processes flatten the cooled electron spectrum leading to a flattening of the synchrotron spectrum, particularly at low and intermediate frequencies. Neglecting bremsstrahlung and Coulomb losses therefore leads to systematically steeper radio spectra and can bias interpretations of observed spectral shapes.
In addition, we showed that free–free absorption and thermal free–free emission can substantially modify radio spectra, particularly below ${\sim}1$~GHz, and should therefore also be accounted for when interpreting radio spectra of star‑forming galaxies (Fig.~\ref{fig:spectra_ff_abs_ff_em}).

Taken together, our results show that several widely adopted assumptions in the interpretation of radio synchrotron observations are not generally valid when confronted with time-dependent CR electron evolution in a realistic galactic environment. In particular, advection-only models can mislead the physical interpretation of the vertical distribution of radio emission, radio spectra do not uniquely encode the global CR electron spectrum, and sub-dominant loss processes can play an important role in shaping observable radio diagnostics.
These findings have important implications for the use of radio observations to constrain CR transport and feedback in galaxies. They further highlight the need for physically motivated, time-dependent forward modelling approaches of radio emission in star-forming galaxies. Future work incorporating additional CR transport mechanisms and re-acceleration processes, as well as a more realistic multiphase ISM, and extending the analysis to simulations in a cosmological context, will be essential to fully exploit the diagnostic power of radio continuum observations.

\section*{Acknowledgements}
CP and JW acknowledge support from the European Research Council via the ERC Advanced Grant ``PICOGAL'' (project ID 101019746). JW acknowledges support by the German Science Foundation (DFG) under grant 444932369. RB is supported by the SNSF through the Ambizione Grant PZ00P2\_223532. PG acknowledges financial support from the European Research Council via the ERC Synergy Grant ``ECOGAL'' (project ID 855130). LJ acknowledges support from the Deutsche Forschungsgemeinschaft (DFG, German Research Foundation) as part of the DFG Research Unit FOR5195 – project number 443220636. RW acknowledges funding of a Leibniz Junior Research Group (project number J131/2022). FvdV is supported by a Royal Society University Research Fellowship (URF\textbackslash R1\textbackslash191703 and URF\textbackslash R\textbackslash241005).

\section*{Data Availability}
The data underlying this article will be shared on reasonable request to the corresponding author.



\bibliographystyle{mnras}
\bibliography{literature}



\appendix

\section{FIR-Radio correlation}\label{app:FRC}

\begin{figure}
    \centering
    \includegraphics[]{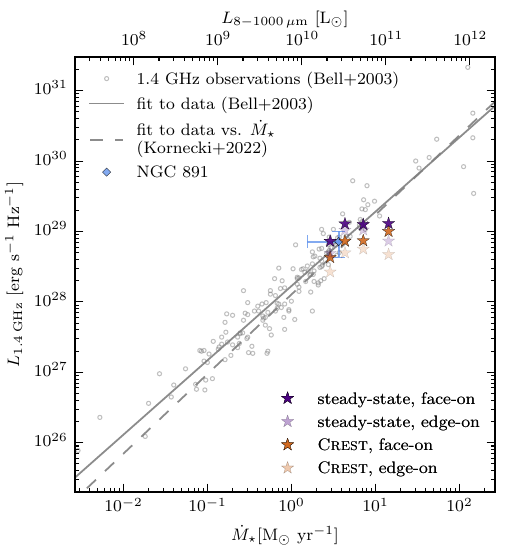}
    \caption{
    Far‑infrared–radio correlation for the simulated galaxy. Both the steady‑state and \crest models are consistent with the observed relation \citep[grey line;][]{2003Bell, 2022Kornecki} when calibrated to the local CR electron‑to‑proton ratio at 10~GeV. Shown are the global 1.4~GHz radio luminosities of the simulated galaxy at four snapshots between $t=1$ and 4~Gyr as a function of SFR (bottom $x$‑axis), with the corresponding FIR luminosity \citep[derived from][]{1998Kennicutt} shown on the top $x$‑axis. 
    The blue diamond shows the observed radio luminosity of NGC~891 (see text for details). 
    Filled symbols show face‑on luminosities, while more transparent symbols indicate edge‑on views.
    The steady-state model without diffusion (purple) adopts an injected electron-to-proton ratio of 0.049, corresponding to an injected energy fraction of $\zeta_\mathrm{ep}=22\%$ in the \crest model (orange).}
    \label{fig:FRC}
\end{figure}

Figure~\ref{fig:FRC} shows the correlation between the global 1.4~GHz radio luminosities of the simulated galaxy and its SFR for both modelling approaches at four snapshots between $t=1$ and 4~Gyr\footnote{Because the SFR declines over time in the simulation, earlier snapshots (e.g. $t=1$~Gyr) appear at higher SFR values (to the right), while later snapshots move progressively toward lower SFR values (to the left).}. Both the time‑dependent \crest modelling and the steady‑state approach with \crayon reproduce the observed relation \citep{2003Bell} at late times ($t\gtrsim2$~Gyr) within the observed scatter.
For NGC~891 (blue data point), the radio luminosity at 1.4~GHz is obtained by extrapolating the observed 1.5~GHz flux density using the measured GHz-frequency spectral index between 1.5 and 6~GHz \citep{2019Schmidt}. The difference between 1.4 and 1.5~GHz is small and does not affect our conclusions. The horizontal error bar reflects the range of published SFR estimates ($1.55~\mathrm{M_\odot\,yr^{-1}}$ from \citealt{2015Wiegert}, $1.88~\mathrm{M_\odot\,yr^{-1}}$ from \citealt{2019Vargas}, and $3.3~\mathrm{M_\odot\,yr^{-1}}$ from \citealt{2012Krause}), while the marker indicates the value derived from the observed FIR-luminosity (top $x$-axis, from \citealt{2021Yoon}), yielding $3.7~\mathrm{M_\odot\,yr^{-1}}$ based on the \citet{1998Kennicutt} relation\footnote{We use the conversion $ \dot{M}_\star/(\mathrm{M_{\odot}\,yr^{-1}})= \epsilon 1.7\times 10^{-10} L_{\mathrm{FIR}}/L_{\odot}$, where the parameter $\epsilon=0.79$ accounts for a \citet{2003Chabrier} IMF \citep{2010Crain}.}. The vertical error bar includes both the flux measurement uncertainty and an additional 20 percent uncertainty on the distance, propagated to the luminosity and added in quadrature.

At $t=1$~Gyr, the simulated radio luminosities lie somewhat below the observed relation in both models. At this early stage, the SFR is still high following the initial starburst, leading to a large photon energy density that exceeds the magnetic energy density when averaged over the galactic disc. While the magnetic field strength has largely saturated by $t=1$~Gyr and remains approximately constant thereafter, the photon energy density decreases with time as the SFR declines. As a result, IC losses dominate over synchrotron losses at early times, suppressing the radio emission relative to later snapshots (e.g.\ $t\gtrsim2$~Gyr), where the magnetic field strength is similar but the photon energy density has dropped. 

Finally, we find that the \crest modelling yields systematically lower radio luminosities than the steady‑state modelling with \crayon, despite both approaches being calibrated to reproduce the locally observed CR electron‑to‑proton ratio in the Milky Way.
This behaviour is another consequence of the fact that radio emission does not directly trace the spatially averaged CR electron spectra (see Section~\ref{Sec:GlobalRadioGlobalElectronSpectra}).

\section{Profiles at different times}\label{app:ProfilesDifferentTimes}

\begin{figure*}
    \centering
    \includegraphics[]{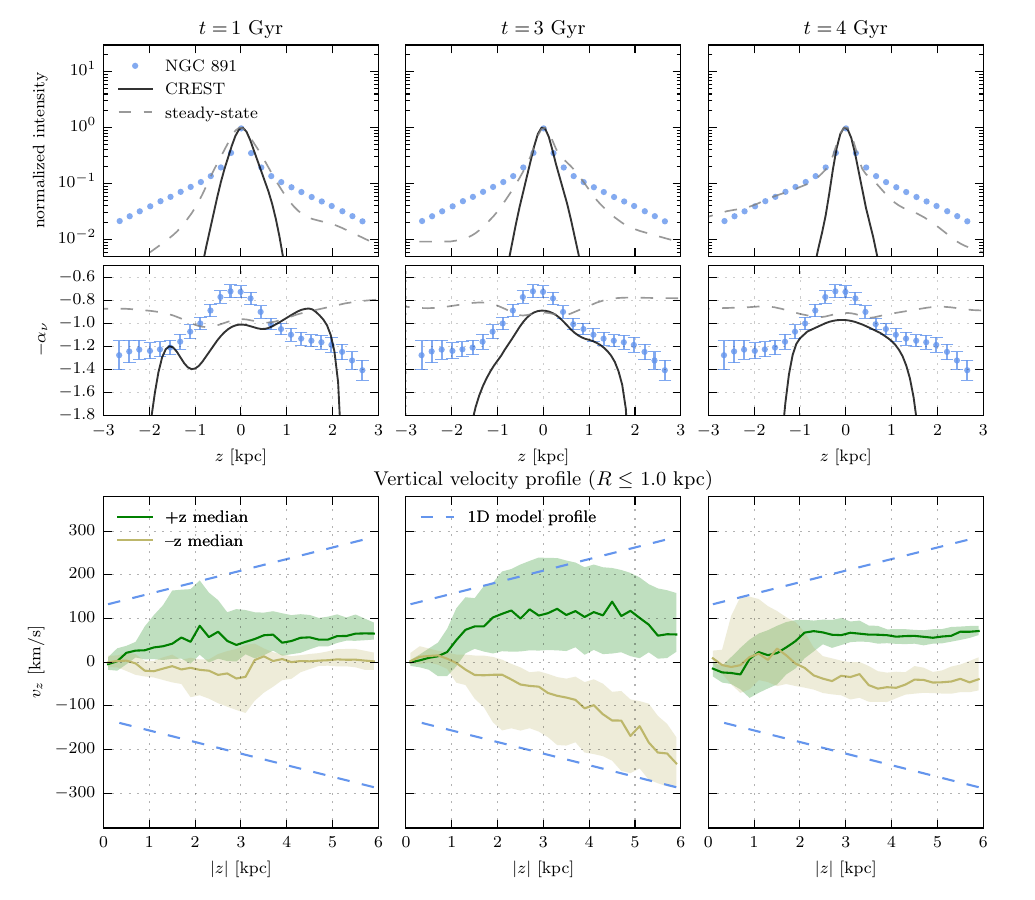}
    \caption{Same as Figs.~\ref{fig:profiles} and \ref{fig:vertical_velocity_profile} but shown at different simulation times.}
    \label{fig:profiles_snaps}
\end{figure*}

The upper two rows of Fig.~\ref{fig:profiles_snaps} show the vertical radio intensity and spectral index profiles, analogous to Fig.~\ref{fig:profiles}, but evaluated at three different times. The \crest intensity profiles exhibit little temporal variation and remain strongly concentrated toward the disc at all times. In contrast, the steady-state profiles obtained with \crayon show a larger degree of time variability and, at some epochs, more closely reproduce the observed vertical extent of the radio emission of NGC~891, particularly below the midplane ($z<0$) at $t=4$~Gyr. 

The temporal behaviour of the spectral index profiles shows the opposite trend. The steady-state models yield nearly time-independent spectral index profiles, whereas the \crest profiles vary strongly with time. In particular, they often lack the approximate symmetry about the midplane that is observed in NGC~891. We note, however, that asymmetric intensity and spectral index profiles are also observed in other galaxies \citep[see e.g. Appendix D in ][]{2018Heesen}.

The lower row of Fig.~\ref{fig:profiles_snaps} shows the corresponding vertical velocity profiles above and below the galactic midplane. These profiles differ between the two sides of the disc and exhibit significant temporal variability, reflecting the time‑dependent nature of the simulated galactic outflows.

\section{Effect of bremsstrahlung losses}

\begin{figure*}
    \centering
    \includegraphics[]{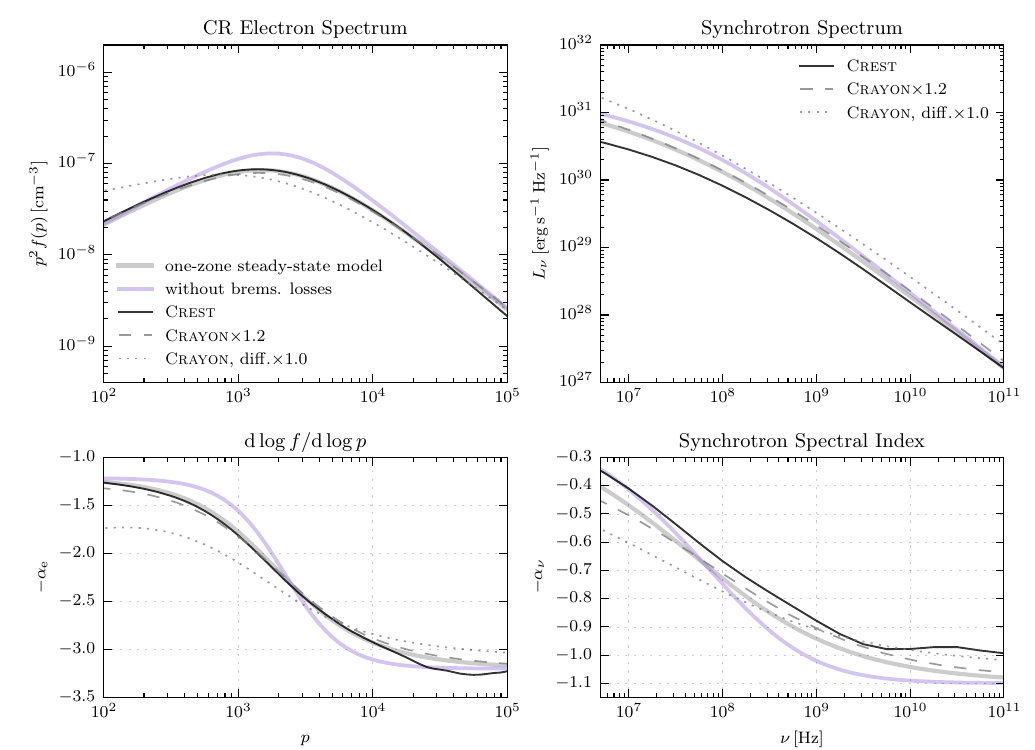}
    \caption{
    Same as Fig.~\ref{fig:electron_synchrotron_spectra}, but showing a one-zone steady-state model in which only bremsstrahlung losses are neglected (purple), while Coulomb losses are retained.}
    \label{fig:electron_synchrotron_spectra_noBrems}
\end{figure*}

In Fig.~\ref{fig:electron_synchrotron_spectra}, we demonstrated the impact of neglecting both bremsstrahlung and Coulomb losses on the shape of the steady-state CR electron and synchrotron spectra. To isolate the role of bremsstrahlung losses, Fig.~\ref{fig:electron_synchrotron_spectra_noBrems} shows the same comparison but with only bremsstrahlung losses omitted.

Comparing the two figures demonstrates that Coulomb losses primarily set the low-energy turnover (around $p\sim2\times 10^3$) of the electron spectrum, while bremsstrahlung losses further modify the spectral shape in the vicinity of this turnover.

\bsp	
\label{lastpage}
\end{document}